\documentclass[]{aa}
\usepackage{graphicx}
\usepackage{natbib}
\usepackage{amssymb}

\begin{document}

\title{Simulating radiation and kinetic processes in relativistic plasmas}
\titlerunning{Simulating relativistic plasmas}
\author{R. Belmont\inst{1}\thanks{\email{belmont@cesr.fr}}, J. Malzac\inst{1}, A. Marcowith\inst{2}}
\institute{Centre d'Etude Spatiale des Rayonnements (OMP; UPS; CNRS), 9 avenue du Colonel Roche, BP44346, 31028, Toulouse Cedex 4, France \and
 Laboratoire de Physique Th\'eorique et d'Astroparticules, IN2P3/CNRS, Universit\'e MontpellierII, CC 70, place Eug\`ene Bataillon, F-34095 Montpellier Cedex 5, France}
\date{Received --- / Accepted ---}

\abstract{Modelling the emission properties of compact high energy sources such as X-ray binaries, AGN or $\gamma$-ray bursts represents a complex problem. Contributions of numerous processes participate non linearly to produce the observed spectra: particle-particle, particle-photon and particle-wave interactions. Numerical simulations have been widely used to address the key properties of the high energy plasmas present in these sources.}
{We present a code designed to investigate these questions. It includes most of the relevant processes required to simulate the emission of high energy sources. }
{This code solves the time-dependent kinetic equations for homogeneous, isotropic distributions of photons, electrons, and positrons. We do not assume that the distribution has any particular shape. We consider the effects of synchrotron self-absorbed radiation, Compton scattering, pair production/annihilation, e-e and e-p Coulomb collisions, e-p bremsstrahlung radiation and some prescriptions for additional particle heating and acceleration.}
{We illustrate the code's computational capabilities by presenting comparisons with earlier works and some examples.  Previous results are reproduced qualitatively but some differences are often found in the details of the particle distribution. As a first application of the code, we investigate acceleration by second order Fermi-like processes and find that the energy threshold for acceleration has a crucial influence on the particle distribution and the emitted spectrum.}{}

\maketitle
\section{Introduction}
High energy sources, such as X-ray binaries, active galactic nuclei (AGN hereafter), or $\gamma$-ray bursts, exhibit spectra detectable to very high energy. This radiation must originate in a plasma for which a significant fraction of the particles have relativistic energies. Understanding the properties of these hot plasmas remains a challenge in the modelling of X- and $\gamma$-ray sources. 

Among the many processes at work, there are particle-particle interactions, such as Coulomb collisions, particle-photon interactions such as Compton scattering, synchrotron radiation, bremsstrahlung emission, or pair production/annihilation, and particle-wave interactions that lead to particle acceleration. However, the way they add or compete is highly non-linear and the cross sections involved are complex. Investigating a large parameter space is required, and in spite of important breakthroughs, these plasmas are still poorly understood. Analytical studies provided interesting qualitative results with approximations, but a more general approach based on numerical simulations is required to explain the details and complexity of contemporary observations.  The first detailed investigations were analytical attempts to model the Compton scattering in thermal plasmas of fixed temperature \citep[][]{BZS71,Sunyaev80,Guilbert81,Zdziarski85,Guilbert86}. In parallel, some of these results were confirmed by Monte-Carlo simulations \citep[e.g.][]{PSS83,Gorecki84}. The additional role of pair production and annihilation in thermal plasmas, whose temperatures were determined self-consistently, was then studied both analytically \citep{Svensson82b,Svensson83,Guilbert85,Kusunose87} and numerically \citep{Zdziarski84,Zdziarski85}. These works constituted significant advances because they explicitly accounted for the back reaction of the radiation field on the plasma temperature. However, they were limited to thermal distributions of particles, whereas significant evidence of strongly non-thermal populations was found in many sources. For instance, spectra of blazars or radio loud AGN were shown to be shaped at least by the synchrotron self-compton emission of purely non-thermal electrons \citep[e.g.][]{Ghisellini98}. At these high energies, accelerated particles cool on very short timescales before they can be thermalized for instance by two body collisions. The balance between this cooling and acceleration typically produces non-thermal distributions. Acceleration processes are still poorly understood. A simple way to simulate the effect of particle acceleration is to inject particles at high energy. Although it does not reproduce exactly the physics involved, this prescription has been widely used and produced interesting results (as shown in most of the references cited here). Significant effort has also been taken in developing more precise modelling of acceleration mechanisms, but in such studies, the radiation field is treated crudely \citep{LKL96,DML96,Li97,Katar06}. 

With the increasing number of considered processes and the increasing precision of their description, numerical analysis has become a prime method of investigation. Even so, a full treatment of the problem accounting for the coupled evolution of inhomogeneous, anisotropic  distributions of leptons and photons, both in momentum and position spaces, appears to be still beyond the capabilities of present-day computers.  Numerical simulations of high energy plasmas have been performed mainly following three different approaches which all make trade-offs  between the various aspects of the problem.

First, the Monte Carlo technique \citep{PSS77,Stern95} allows one to follow particles and photons in space, time, and energy as they undergo mutual interactions. It solves the full radiative transfer problem and accounts explicitly for geometrical effects. At present, the MC method is probably the mos effective way to model fully 3-dimensional problems. However, this detailed procedure is time consuming, particularly when modelling the rapid dynamics of the non-thermal electron population in momentum space \citep{mj00}, and when synchrotron self-absorption effects are important \citep[see discussion in][]{Stern95}. For this reason, the Monte-Carlo methods have been applied to date to pure Maxwellian plasmas and/or steady state problems with 3D geometry.


Another method that accounts correctly for the geometry, involves solving numerically the exact radiation transfer equation for given geometries and particle distributions \citep{Poutanen96}. This method is far more efficient than Monte Carlo simulations which makes it easier to compare with data. It is, however, far less versatile than Monte Carlo methods and does not solve the kinetic equations for particles. The back reaction of the radiation field on the particle distribution is modelled only for the assumption of a Maxwellian plasma (in which case the plasma temperature may be adjusted according to energy balance). The method applicability is also limited to the resolution of steady state problems. 

The third approach, which we adopt in this paper, abandons the detailed description of the geometry to concentrate on the kinetic effects. It consists of solving the local kinetic equations for the particle and photon distributions.To maximaze efficiency, radiative transfer is usually modelled with a simple photon escape probability formalism assuming isotropic photon and particle distributions. This method can be applied to different, possibly time-dependent, problems for which geometry does not play a crucial role\footnote{For problems in which geometry is important, radiative transfer can in principle be accounted for by coupling this kinetic code with a radiation transfer solver or a Monte Carlo code  (as demonstrated by \cite{BL01}), although  computing time may then become a serious issue.}. Within the limits of the one-zone approximation, it is more efficient than other methods and allows for fast data fitting. 


The first detailed investigations of high energy plasmas with this technique concentrated on thermal pair plasmas \citep{FBGPC86,Ghisellini87}. More precise modelling was then proposed in which the particle distributions were decomposed into the sum of a thermal low-energy pool and an arbitrary high energy tail \citep{LZ87,Svensson87,Coppi92,ZLM93,GHF93,LKL96}. The latter models have been applied most to fitting and interpreting data. They do not however describe the possible deviation from a Maxwellian distribution at low energy, nor do they address explicitly any thermalization process.  Only recent numerical work considered fully arbitrary distributions of particles. \citet{GHS98} concentrated on the role of synchrotron self-absorbed radiation in AGN. They confirmed previous analytical results \citep{GGS88} by demonstrating that the exchange of energy between particles by means of synchrotron photons can be an efficient thermalization process in magnetized sources. These simulations focused however on this specific interaction, and other processes were only considered by crude approximations, particularly Compton radiation, or not considered at all. \citet{NM98} investigated the thermalization of arbitrary distributions by two-body particle interactions and heating by high energy protons. The additional role of synchrotron radiation was not however considered. The most complete numerical treatment of high energy plasmas was probably one developed in the context of $\gamma$-ray bursts by \citet{PW05}. Our code is similar to this study but differs in that these authors considered neither particle stochastic acceleration nor the effect of Coulomb losses

The code presented here solves the time-dependent equations simultaneously for isotropic, arbitrary photon, electron, and positron distributions. The evolution of these populations is modelled in time while being affected by self-absorbed cyclo-synchrotron radiation, Compton scattering, pair production/annihilation, e-e and e-p Coulomb collisions, self-absorbed e-p bremsstrahlung radiation, and additional particle acceleration and heating. Each process is described with minimal approximations and by using in most of cases the exact cross sections. For instance, the formulae used for the synchrotron emission and absorption are valid from the sub-relativistic to the ultra-relativistic regime. This numerical strategy allows one to investigate many different astrophysical situations that occur in various high energy sources. 

The structure of this paper is as follows. Section~\ref{sec:rad} provides a description of the microphysics adapted into our code. Then, in Sec.~\ref{sec:numsol}, we present the numerical techniques. Finally, in Sec.~\ref{sec:testsandapps} the code is tested against previous published results, providing an overview of its capabilities.

\section{Radiation and kinetic processes}\label{sec:rad}
We describe the processes included in the code. We present first the general notation used in this paper. The particle energy is described either by the relativistic Lorentz factor $\gamma=E/m_ec^2$, by the adimensional momentum $p={\cal P}/m_ec=\sqrt{\gamma^2-1}$, or by the beta parameter $\beta=p/\gamma$, where $m_e$ is the electron mass and $c$ is the speed of light. Similarly, the photon energies are described by their frequencies $\nu$ or $\omega=h\nu/m_ec^2$. The particle and photon populations are described by their angle-integrated distribution functions $N_{e^\pm} = \frac{\partial {\cal N}_{e^\pm} }{\partial^3 x \partial p} $ and $N_\nu = \frac{\partial {\cal N}_\nu }{\partial^3 x \partial \nu} $, where $\partial {\cal N}_{e^\pm}$  and $\partial {\cal N}_{\nu}$ are the number of electrons, positrons and photons per unit volume $\partial^3 x$ and per unit momentum $\partial p$ or frequency $\partial \nu$. For simplicity, the total lepton distribution is also used: $N_e=N_{e^-}+N_{e^+}$.
Finally, $R$ is the typical length scale of the emission region. Since we consider a homogeneous medium, most of the processes are scale-free, meaning that most quantities are simply proportional to $R$, $R^2$, or $R^3$. For those quantities, $R$ determines only the overall normalization factor. For instance, the total luminosity of unmagnetized sources scales as $R^3$, but there is no reference scale in the problem. The only process that explicitly involves a reference length scale is the synchrotron self-absorption, since it is independently determined by both the magnetic intensity and the total magnetic energy of the source, which is related directly to the source size for a given magnetic intensity. 

\subsection{Self-absorbed Synchrotron Radiation}
The cyclo-synchrotron radiation process is produced by charged particles gyrating along magnetic field lines. It is one of the most important processes in astrophysics and is invoked to explain the radio emission of many magnetized sources, such as supernova remnants, pulsar wind nebulae, AGN, X-ray binaries, and $\gamma$-ray bursts. In particular, it produces soft photons that can be up-scattered by Compton scattering producing the well known double humped synchrotron self-compton spectra used to model for example the emission of blazars or radio-loud AGN. The reverse process, cyclo-synchrotron absorption, is less well known, although efficient at low energy. Besides explaining some observed spectra, cyclo-synchrotron emission and absorption both influence the involved particles by cooling and heating them respectively. As we discuss later, these interactions can thermalize high energy particles \citep[the so-called {\it synchrotron boiler},][]{GGS88}. 

Following the main assumption of the code, we assume isotropy in the radiation field and particle pitch angle. These are accurate approximations when the magnetic field is tangled. The cyclo-synchrotron emission and absorption are characterized by the emissivity spectrum $j_s(p,\nu)$ (erg s$^{-1}$Hz$^{-1}$) of one single particle of momentum $p$ and the cross section $\sigma_s(p,\nu)$ (cm$^2$). Both quantities are related to each other by the formula\footnote{The term $1/4\pi$ in Eq. \ref{Eq:sigmas} and \ref{eq_sync3} relates to the average over the solid angle} \citep{Leroux61,GGS88,GS91}:
\begin{equation}
\label{Eq:sigmas}
\sigma_s(p,\nu) = \frac{1}{4\pi} \frac{1}{2m_e\nu^2} \frac{1}{p^2} \partial_p \left[ p\gamma j_s(p,\nu) \right] \label{sigmas}
\end{equation} 
where $m_e$ is the electron mass. The emissivity and cross section depend on the magnetic field, whose intensity is characterized by the {\it magnetic compactness} \citep{GGS88}:
 \begin{equation}
 l_B = \frac{\sigma_TR}{m_ec^2} \frac{B^2}{8\pi} ~.
 \end{equation}
As mentioned earlier, the synchrotron self-absorption depends explicitly on the source size. Although the overall normalization of the emissivity and absorption is only proportional to a combination of the magnetic field intensity and source size (namely $l_B$), their shape depends on the cyclotron frequency $\nu_B = eB/2\pi m_e c$, which depends only on the magnetic field intensity.  For a given magnetic compactness parameter, simulations of sources with different sizes correspond to cases with different magnetic field intensities and therefore produce different observed spectra.

In uniform systems, the time evolution in the mean intensity integrated over solid angles $I_\nu=h\nu c N_\nu$ (erg s$^{-1}$ Hz$^{-1}$ cm$^{-2}$) is described by the equation: 
\begin{equation}
\partial_t I_\nu/c = \mu_\nu- \kappa_\nu I_\nu \label{eq_Inu}
\end{equation}
with 
\begin{eqnarray}
\mu_\nu &=& \int N_e j_s(p,\nu) dp \quad \quad \quad (\mbox{erg cm}^{-3}\mbox{s}^{-1}\mbox{Hz}^{-1}) ~, \label{eq_In2}\\
\kappa_\nu &=& \int N_{e} \sigma_s(p,\nu) dp \quad \quad \quad (\mbox{cm}^{-1})~.\label{eq_In3}
\end{eqnarray}
We note that this equation differs from one often used in previous works \citep[for example][]{GGS88,GHS98}. These have concentrated mostly on the steady state properties of magnetized sources and did not include this time dependence. To account for photon escape and the non-absorbed part of the observed spectra, a finite size domain is assumed in these papers and the space equation $\partial_x I_\nu = \mu_\nu- \kappa_\nu I_\nu$ is solved for uniform emissivity $\mu_s$ and absorption $\kappa_s$ on a typical length scale $R$, which yieds the synchrotron self-absorbed radiation: $I_\nu = \mu/\kappa(1-e^{-\kappa R})$. This is an approximate way to deal with the space dependence of the simulated system since it implies a non-uniform synchrotron radiation whose feedback onto the lepton equation would need to be incorporated in a fully space-dependent model. This solution would holds also only when synchrotron self-absorption is involved. Other processes, such as Compton scattering or pair production/annihilation, would also contribute to the emissivity $\mu$ and absorption $\kappa$ in some way and in this description it is unclear how the synchrotron interaction couples with others in the global equation for the radiation field. We consider a homogeneous (or averaged) radiation field which, associated with the photon escape probability (see discussion in Sect. \ref{injesc}) represents another approximate way to model crudely the geometry. However, this method solves the exact time-dependent equation and the synchrotron emission is added consistently to other emission processes.

Simultaneously, the equation for the time evolution of the lepton populations is \citep{McCray69,GGS88}:
\begin{equation}
\partial_t N_{\rm{e^\pm}} = \partial_p \left( \frac{\gamma}{p}A^{\rm{s}}_e N_{\rm{e^\pm}} \right) + \frac{1}{2}\partial_{p}\left[ \frac{\gamma}{p} \partial_p \left( \frac{\gamma}{p}  D^{\rm{s}}_e N_{\rm{e^\pm}} \right) \right]\label{eq_sync}
\end{equation}
where the $\gamma/p$ factors result from the choice of $p$ as a variable instead of $\gamma$ and 
\begin{eqnarray}
A^{\rm{s}}_e &=& \frac{1}{mc^2} \int \left( j_s-\sigma_s I_\nu \right) d\nu \quad\quad\quad \rm{(s}^{-1}\rm{)} \label{eq_sync2} ~, \\
D^{\rm{s}}_e &=& \frac{1}{4\pi} \frac{1}{m_e^2c^2} \int \frac{j_sI_\nu}{\nu^2} d\nu  ~\quad\quad\quad\quad \rm{(s}^{-1}\rm{)} \label{eq_sync3} ~.
\end{eqnarray}
Equation \ref{eq_sync} can be written in many different ways that are analytically equivalent. The one used (Eq. \ref{eq_sync}-\ref{eq_sync3}), associated with a specific numerical scheme to estimate derivatives, enable good numerical accuracy. In particular, the energy conservation can be easily satisfied to machine precision when particles and the photon field exchange energy, since $dE_e/dt = mc^2 \int \gamma\partial_p\left[ A_e N_e\right] dp = - \int\hspace{-.15cm}\int (j_s-\sigma_sI_\nu) N_e dpd\nu =-\int (\mu_s-\kappa_s I\nu) d\nu = -dE_\nu/dt$.

There is no exact analytical expression for pitch angle averaged- and photon direction integrated- emissivity and absorption that is valid in all regimes. The exact values are derived from numerical integrations, which are time-consuming and hard to perform, especially for low energy particles when the emission is dominated by a few narrow harmonics. However, approximations have been proposed, which are valid in some regimes \citep[e.g.][]{Marcowith03}. We use a combination of two approximations. For sub-relativistic particles, we use the formula for $j_s$ first proposed by \citet{GHS98} and corrected by \citet{KGSG06} both to match the relativistic spectrum more accurately and to describe the spectrum more closely close to the minimal frequency. This approximation is less accurate for very low energy particles (typically $\beta \lesssim 0.1$). However, for most astrophysical cases, the emission is produced mainly by energetic particles, and this regime has little influence on the total particle distribution and radiation field. Numerical experiments have confirmed that the choice of $j_s$ and $\sigma_s$ at low energy has negligible effect. In the relativistic regime, we use the well-known synchrotron power spectrum integrated over an isotropic distribution of pitch angles \citep{CS86,GGS88}. We note that we correct this formula similarly to match more accurately the sub-relativistic regime. The transition between both regimes is achieved by applying an exponential threshold/cut-off to the formulae at $\gamma= 2$. The formulae for $\sigma_s$ are computed analytically from $j_s$ with Eq. \ref{sigmas} and implemented into the code.

\subsection{Compton Scattering}

Compton scattering is a well-known interaction between leptons and photons. Most of the previous studies assume thermal distributions of particles. In such cases, only the temperature and total number of particles are computed. This assumption enables rapid computation with simple formulae but omits the physics of non-thermal particles. We therefore use exact analytical expressions for unpolarized radiation and arbitrary distributions of isotropic particles and photons.

\subsubsection{Basic equations}

The effect of Compton scattering can be described by the sum of individual encounters over the entire distributions. The scattering of isotropic photons of energy $ h\nu_0 = \omega_0 m_ec^2$ off isotropic particles of energy $E_0= \gamma_0 mc^2$ is fully characterized by the resulting distribution of scattered photons $\sigma_{\rm{c}}(p_0,\nu_0 \rightarrow \nu)$. This differential Klein-Nishina cross section has been computed by several authors \citep{Jones68, Brinkmann84,NP94}. The numerical evaluation of these analytical expressions can be difficult, in particular for low or high energy particles and photons. We use an expression based on the formulae by \citet{Jones68} and modified to overcome numerical accuracy issues \citep{Belmont08}.

The exact time evolutions of the full particle and photon distributions are described by the following equations:
\begin{eqnarray}
\partial_t N_{e^\pm}(p) &=&  \int\hspace{-.3cm} \int N_{e^\pm}(p_0) N_\nu(\nu_0) c \sigma_{\rm{c}}(p_0,\nu_0\rightarrow \nu(p))  dp_0d\nu_0  \nonumber  \\ 
                                                    & & \quad \quad  \quad -  N_{e^\pm}(p) \int N_\nu(\nu_0) c \sigma^{\rm{c}}_0(\nu_0,p) d\nu_0 \label{compt_e_int} ~, \\
\partial_t N_\nu(\nu) &=&  \int \hspace{-.3cm} \int N_e(p_0) N_\nu(\nu_0) c \sigma_{\rm{c}}(p_0,\nu_0\rightarrow \nu)  dp_0d\nu_0  \nonumber \\ 
                                                    & & \quad \quad  \quad -  N_\nu(\nu) \int N_e(p_0) c \sigma^{\rm{c}}_0(\nu,p_0) dp_0 ~,  \label{compt_nu_int} 
\end{eqnarray}
where the photon frequency $\nu(p)$ is constrained by the energy conservation during one scattering event: $h\nu(p)-h\nu_0 + \gamma m_ec^2-\gamma_0 m_ec^2=0$. For each distribution, the first integral provides the number density of scattered particles/photons that have a particular energy after one single scattering and the second one is the probability that particles/photons of this energy are scattered to some other energy. This is what we refer to as the {\it integral approach}. As discussed hereafter, the numerical computation of this integral suffers from accuracy issues because of discretization.

In the small-angle scattering limit, when the scattered photons (or particles) have energies similar to those incoming, a {\it Fokker-Planck approximation} can be used (FP hereafter). In this case, a second-order series expansion of the exact equations produces the FP evolution equations for the different species:
\begin{eqnarray}
\partial_t N_\nu(\nu) &=&  \partial_\omega \left[A^{\rm{c}}_\nu N_\nu\right] + \frac{1}{2} \partial^2_{\omega^2}\left[D^{\rm{c}}_\nu N_\nu\right]  ~,\\
\partial_t N_{e^\pm}(p) &=&  \partial_p \left(\frac{\gamma}{p}A^{\rm{c}}_e N_{e^\pm} \right) + \frac{1}{2} \partial_{p}\left[ \frac{\gamma}{p} \partial_p\left(\frac{\gamma}{p}D^{\rm{c}}_e N_{e^\pm}\right) \right] \label{eq_FP} 
\end{eqnarray}
with 
\begin{eqnarray}
A^{\rm{c}}_\nu = - \int N_e c\sigma_1^{\rm{c}}(p,\nu) dp &,\quad& D^{\rm{c}}_\nu = \int N_e c\sigma_2^{\rm{c}}(p,\nu) dp ~, \label{Ac_nu} \\
A^{\rm{c}}_e = \quad \int N_\nu c\sigma_1^{\rm{c}}(p,\nu) d\nu   &,\quad& D^{\rm{c}}_e = \int N_\nu c\sigma_2^{\rm{c}}(p,\nu) d\nu ~, \label{Ac_e}
\end{eqnarray}
and where we have introduced the first 3 moments of the scattered photon distribution: 
\begin{eqnarray}
\sigma_{0}^{\rm{c}}(p_0,\nu_0) &=& \int \sigma_{\rm{c}}(p_0,\nu_0 \rightarrow \nu) d\nu ~, \\
\sigma_1^{\rm{c}}(p_0,\nu_0) &=& \int (\omega-\omega_0) \sigma_{\rm{c}}(p_0,\nu_0 \rightarrow \nu) d\nu ~,\\
\sigma_2^{\rm{c}}(p_0,\nu_0) &=& \int (\omega-\omega_0)^2 \sigma_{\rm{c}}(p_0,\nu_0 \rightarrow \nu) d\nu ~,
\end{eqnarray}
which are the total cross section, the mean photon energy, and the dispersion, respectively.

This approximation allows far quicker computation since, when the first moments have been tabulated, only single integrals are required whereas the exact computation requires double integrals.  However, the Fokker-Planck approach used to model the evolution of particle and photon distributions is valid only in regions of the incident energy space $(\nu_0,p_0)$ for which the relative energy exchange in one scattering is small:  $\Delta p (p_0,\nu_0) /p_0 << 1$ and $\Delta \nu (p_0,\nu_0) / \nu(\nu_0) << 1$, respectively. These conditions will be presented in a forthcoming publication \citep{Belmont08}.

\subsubsection{Numerical strategy}
In contrast to the Fokker-Planck approximation, the integral approach is exact analytically. However, when used to compute the evolution numerically, it leads to some numerical issues directly related to the use of non-linear grids \citep{NM98}. 

With logarithmic grids, the energy bin size is larger at high energy. For example when, low energy photons are up-scattered from high energy particles, their relative energy gain is high, and these photons are scattered numerically from low energy bins to higher energy bins. During this interaction, the particles lose only a small fraction of their energy. If the energy bin size is too large, these particles remain in their original bin and numerically, they do not lose energy. The energy balance is not therefore exactly satisfied and the error can propagate and become large when the density of low energy photons is also high. Although less relevant to most astrophysical situations, a symmetrical problem appears when high energy photons are scattered by low energy particles. 

This numerical issue is not present in regions of the incident energy space $(p_0,\nu_0)$ for which the scattered distributions are far wider than the energy bin size: $\Delta \nu (p_0,\nu_0) /\delta\nu(\nu_0) >> 1$ and $\Delta p (p_0,\nu_0) /p_0 >> 1$ for the evolution of the photon and particle distributions respectively. After selecting the ranges and resolution of the photon and particle energy grids, these conditions constrain the region in which the integral approach is valid. 
Fortunately, the regions for which the integral and the Fokker-Planck approaches are valid are in part complementary. The code therefore combines the two approaches:\\
$\bullet$ For the particle evolution: \\
In the integral approach, the integration over the photon distribution in Eq. \ref{compt_e_int} is only completed above a given photon energy $\nu_c(p_0)$ that depends on the incident particle energy, whereas the integrals on frequency in Eq. \ref{Ac_e} are completed up to $\nu_c$ in the Fokker-Planck approach. The total time evolution is then given by the sum of both contributions: $\partial_t N_{e^\pm} = \left(\partial_t N_{e^\pm}\right)_{\rm{FP}} + \left(\partial_t N_{e^\pm}\right)_{\rm{Integral}}$.  \\
$\bullet$ For the photon evolution: \\
A similar combination is used for the photon equation, with the definition of a critical particle momentum $p_c(\nu_0)$ under which the FP approach is used and above which the integral approach is used. 

If the number of energy bins per decade is too small, the validity domains for both calculations may become independent and the accuracy of the computation may decrease. This will be true, however, only for small regions of the grids for which there are few particles and photons, corresponding to very small errors.  By combining the two approaches, we find empirically that an energy resolution of typically 10 energy bins per decade provides errors that are smaller than other truncation errors.

\subsection{Pair production/annihilaton}

As for Compton scattering, we describe the pair production and annihilation for the case of isotropic distributions of particles and photons. Single photon-photon pair production and pair annihilation events are characterized by the differential cross sections $\sigma_{\rm p}(\nu_1,\nu_2\rightarrow p)$ which corresponds to the pair (i.e. electrons or positrons) momentum spectrum produced by the recombination of photons of frequencies $\nu_1$ and $\nu_2$, and $\sigma_{\rm{a}}(p_-,p_+\rightarrow\nu)$, which corresponds to the emission spectrum generated by the annihilation of one electron of momentum $p_-$ and one positron of momentum $p_+$, respectively. Then, the evolution of the distributions is described by:
\begin{eqnarray}
\partial_t N_{e^\pm}(p) &=& \int \hspace{-.3cm}\int N_\nu(\nu_1) N_\nu(\nu_2)c \sigma_{\rm{p}}(\nu_1,\nu_2 \rightarrow p ) d\nu_1 d\nu_2 \nonumber \\
                                           & & \quad \quad- N_{e^\pm}(p) \int N_{e^\mp}(p') \sigma_0^{\rm{a}}(p,p') dp' ~,\\
\partial_t N_{\nu}(\nu) &=& \int \hspace{-.3cm}\int N_{e^-}(p_-) N_{e^+}(p_+)c \sigma_{\rm{a}}(p_-,p_+ \rightarrow \nu ) dp_- dp_+ \nonumber \\
                                           & & \quad \quad- N_\nu(\nu) \int N_\nu(\nu') \sigma_0^{\rm{p}}(\nu,\nu') d\nu' 
\end{eqnarray}
where, as for Compton scattering, the zeroth moment of both the annihilation spectrum $\sigma^{\rm{a}}_0(p_-,p_+) =1/2\int \sigma_{\rm{a}}(p_-,p_+\rightarrow \nu ) d\nu$ and the pair-produced distribution  $\sigma^{\rm{p}}_0(\nu_1,\nu_2) = 2\int \sigma_{\rm{p}}(\nu_1,\nu_2\rightarrow p ) dp$ have been used\footnote{The $1/2$ and $2$ factors result from the fact that one pair annihilation produces 2 photons and one photon-photon annihilation produces 2 leptons.}. The analytical expressions for photon-photon pair production and pair annihilation correspond to Eqs. (24-29) of \citet{BS97} and Eqs. (23,33,55-58) of \citet{Svensson82a}, respectively. 

In contrast to Compton scattering, there are no numerical problems in computing directly the integral over the particle and photon distributions, even for low resolution grids.

\subsection{Coulomb scattering}
Two kinds of Coulomb-type interactions are considered: scattering of leptons by other leptons and scattering of leptons by protons. 
When the particles are not too energetic, e-e collisions tend to thermalize the pair distributions. In some astrophysical situations, protons are assumed to have a high temperature, so that e-p collisions tend to heat the lepton populations.  
Both kinds of interactions are described by the Boltzmann collision integral. Computing this integral numerically is challenging, mainly because the Coulomb cross section diverges when the energy exchange becomes too small. We assume instead the approximation of small angle scattering, which leads to simple Fokker-Planck equations. As has been already discussed in the literature, the contribution of large angle scattering is negligible \citep{DL89,NM98}. 

\subsubsection{Moeller and Bhabha e-e scattering}
The FP coefficients ($A^{e-e}$ and $D^{e-e}$) for Moeller $e^\pm-e^\mp$ and Bhabha $e^\pm-e^\pm$ scattering are similar. The relative difference is typically of the order of $\sim1/\ln{\Lambda}$, where the Coulomb logarithm $\ln{\Lambda}$ is large \citep{DL89}. Neglecting these terms, the FP coefficients are computed from the following integrations over the mirror distributions:
\begin{eqnarray}
A^{e-e}_{e\pm}(p_\pm) &=& \int N_{e^\mp}(p_\mp) a_e(p_-,p_+) dp_\mp ~,\\
D^{e-e}_{e\pm}(p_\pm) &=& \int N_{e^\mp}(p_\mp) d_e(p_-,p_+) dp_\mp ~.
\end{eqnarray}
The specific coefficients $a_e$ and $d_e$ were first given by \citet{NM98} (Eq. 24 and 35 respectively). The equation for $d$ however contains typos that were corrected by Eq. 6 of \citet{Blasi00}.

\subsubsection{Coulomb e-p scattering}
\label{e-p}
In some astrophysical situations, protons are believed to have temperatures far higher than those of electrons. In these cases, Coulomb collisions with leptons tend to heat the electrons.
As for the e-e collisions, the FP coefficients ($A^{e-p}$ and $D^{e-p}$) for e-p Coulomb scattering are computed from integrals over the proton distribution:
\begin{eqnarray}
A^{e-p}_{e\pm}(p_e) &=& \int N_{p}(p_p) a_p(p_e,p_p) dp_p \label{eq_A_coul} ~,\\
D^{e-p}_{e\pm}(p_e) &=& \int N_{p}(p_p) d_p(p_e,p_p) dp_p \label{eq_D_coul}
\end{eqnarray}
where $a_p$ and $d_p$ are derived from Eqs. (45-48) of \citet{NM98}. The code could calculate the exact proton distribution as for electrons and positrons. However, it would add a fourth kinetic equation and require more computational time. We use instead a thermal proton distribution. Depending on the physical situation being modelled, the proton temperature can be set to equal a constant at the beginning of the simulations or evolved with time to provide a constant electron heating (see section \ref{acc}). 

\subsection{Particle and photon injection}
\label{inj}
The code also allows for injection of particles into the system. This can represent a real injection from an outer source (e.g. particles from the standard disc into an ADAF). But injection of high energy particles is most commonly used to mimic particle acceleration processes. Any distribution $\dot{N}_{e^\pm}^{\rm{inj}}$ can be injected at each time step. Thermal, Gaussian, power-law, and mono-energetic injections have already been included into the code. The injection of particles is controlled by the particle injection compactness\footnote{Note that some authors use instead the kinetic energy to define the compactness parameter: $ l_{e^\pm} = 4\pi R^2\sigma_T/3c  \int (\gamma-1) \dot{N}_{e^\pm}^{\rm{inj}}dp$. }: 
\begin{equation}
l_{e^\pm} = \frac{\dot{E}_{\rm{inj}} }{m_ec^3R/\sigma_T} = \frac{R^2\sigma_T}{c}  \frac{4\pi}{3} \int \gamma \dot{N}_{e^\pm}^{\rm{inj}} dp ~. 
\end{equation}
Similarly, the code allows for photons injection that can account for example for seed photons from the cold accretion disc in X-ray binaries. The photon injection rate is controlled by the parameter: 
\begin{equation}
 l_\nu =\frac{L_{\rm{inj}} }{m_ec^3R/\sigma_T} =\frac{R^2\sigma_T}{c}  \frac{4\pi}{3}  \int \frac{h\nu}{m_ec^2} \dot{N}_{\nu}^{\rm{inj}}d\nu ~.
 \end{equation}
So far the code can account for photon injection with a pure black-body spectrum of specific temperature or a multi-temperature black-body spectrum characterised by the inner and outer temperatures.

\subsection{Particle and photon losses: geometry of the source}
\label{injesc}
In general, photons and particles can also escape from the system. The precise way in which they escape depends on the detailed geometry of the simulated source, which goes beyond the scope of such a 1-zone kinetic code.  Although the losses must occur at the boundaries of the simulated plasma, we use a standard method and we consider all photons (or particles) to have the same averaged probability $p_\nu^{\rm{esc}}$ (or $p_{e^\pm}^{\rm{esc}}$) of escape. We assume spherical geometry and use probability laws describing this geometry approximately. The total luminosity of the source at each frequency is then:
\begin{equation}
L_\nu = \frac{4\pi R^3}{3} h \nu p^{\rm{esc}}_\nu N_\nu
\end{equation}

The code allows for fully trapped pairs in extremely magnetized systems (no loss) and for freely escaping pairs (with an escape probability proportional to their velocity: $p_{e^\pm}^{\rm{esc}} =\beta c/R $). Other escape laws can be defined and implemented easily. 

The photon escape is more debatable. The photon dynamics are affected strongly by Compton scattering. high energy photons do not scatter and can escape freely, whereas, when the optical depth is large, low energy photons can be scattered so significantly that they become trapped in the system. Depending on their energy, the exact way in which they escape strongly involves geometrical effects. 
In the code, we use the escape rate $r_\nu^{\rm{esc}}$ (or escape probability defined as $p^{\rm esc}_\nu=r^{\rm esc}_\nu \times R/c$) derived by \citet{LZ87}: 
\begin{equation}
r^{\rm{esc}}_\nu = 1/T^\nu_{\rm{esc}} = \frac{c/R}{1+\tau(\omega) f(\omega)/3 } \label{pesc}
\end{equation}
where $T_\nu^{\rm{esc}}$ is the averaged escape time,
\begin{equation}
\tau (\omega=h\nu/m_ec^2) = R\sigma_T  \int{N_{e} \frac{\sigma^{\rm{c}}_0(\nu,p)}{\sigma_T} dp}
\end{equation}
is the Compton interaction probability of photons (of frequency $\nu$) with the lepton distributions and
\begin{equation}
f(\omega=h\nu/m_ec^2)=  \left\{ \begin{array}{ll} 1 & \mbox{for } \omega \le 0.1 \\ (1-\omega)/0.9 & \mbox{for } 0.1 < \omega \le 1 \\ 0 & \mbox{for } \omega \ge 1 \end{array} \right. \end{equation}
is a relativistic factor correcting for the fact that forward collisions are less efficient in trapping the photons.

The choice of escape probability is important and different laws can lead to substantially different results\footnote{For instance, when comparing his results with those of \citet{LZ87}, \citet{Coppi92} attributed the difference to different descriptions of the microphysics, whereas, from the results of several simulations, we believe that the difference is due to a different choice for the escape rate: he used $(R/c)/(1+\tau(\omega) f(\omega))$ instead of Eq. \ref{pesc}.}. Although it was shown that this escape probability reproduces well the results of Monte Carlo simulations in a spherical geometry \citep{LZR87,LZ87}, the conclusions of \citet{Stern95} imply that the escape probability may be slightly underestimated. Since the escape luminosity must equal the injected power in a steady state, a smaller escape probability implies a stronger radiation field inside the source. The consequences on the shape of the photon and lepton distributions become significant only when pair production and annihilation are extremely efficient (typically at high optical depths). Figure \ref{Coppi92a} shows spectra in these cases when the deviation from the Monte Carlo simulations become significant. Other escape probabilities were proposed \citep[e.g.][]{Stern95} to reproduce more successfully the results from MC simulations in some specific regimes, but none were shown to be fully consistent with MC results. The use of an escape probability to mimic geometrical effects is of course the main limitation of our code. However, significant deviations appear only in optically thick plasmas when steep gradients in temperature and intensity appear, whereas jets and coronae in XRB and AGN are optically thin media, the largest optical depths observed being $\tau \approx 2-3$. The precise geometry of the sources is also unknown and describing accurately the escape probability in one peculiar geometry is therefore not necessarily helpful.

\subsection{Additional particle heating/acceleration}
\label{acc}

Particle heating and acceleration are probably amongst of the most mysterious problems of high energy sources. Observations show evidence for hot plasmas or high energy tails in the particle distributions, but little is known about the precise mechanisms that generate these populations. Most previous work did not address this problem directly. Non-thermal high energy particles were instead injected into the system with an arbitrary (usually  power law) distribution.  This ad hoc injection assumes an instantaneous acceleration of particles. It does not take into account the fact that particle acceleration has to compete with other cooling processes. Another simple approach, often used to account for lepton heating, consists of assuming that power is provided by some unspecified process to the supposedly thermal distribution of electrons.  
These prescriptions for particle acceleration and heating are implemented in the code. 
However, in addition, we also attempted to follow a more physical approach by implementing two additional specific mechanisms for heating and acceleration, namely Coulomb heating and second order Fermi acceleration. 

$\bullet$ e-p Coulomb-like heating: \\
As has already been discussed, collisions with hot protons can heat the pair distributions. The way in which the interaction is adapted into the code is described in Sect. \ref{e-p}. When the Thomson optical depth is lower than unity, this heating is known to become inefficient and other processes must operate, which are not fully understood. A possible means of accounting for this additional heating is to mimic the heating by thermal protons but with enhanced efficiency \citep{NM98}. Although we do not aim to describe any precise microphysics, this heating prescription estimates consistently both FP coefficients: the heating rate and its related diffusion coefficient.
For this heating prescription, the temperature is set and the total number of protons is constrained by the initial neutrality. The usual cooling and diffusion coefficients $A_{e^\pm}^{e-p}$ (Eq. \ref{eq_A_coul}) and $D_{e^\pm}^{e-p}$ (Eq. \ref{eq_D_coul}) are then multiplied by an efficiency factor $\eta$. This efficiency is computed at each time step so that the total heating is controlled by a constant heating parameter
\begin{equation}
l_c=\frac{4\pi}{3} \frac{R^2\sigma_T}{c} \int -\eta A_{e^\pm}^{e-p} N_e dp ~. \label{lh}
\end{equation}

\noindent $\bullet$ 2nd order Fermi-like acceleration: \\
This type of acceleration could be generated for example by the interaction between the electron and wave turbulence. Diffusion of particles in momentum space is then described by the general equation:
\begin{equation}
\frac{\partial f}{\partial t} = \frac{1}{p^2} \frac{\partial}{\partial p } \left[ p^2 D^{\rm{diff}} \frac{\partial f}{\partial p}  \right]
\end{equation}
where $f(p)$ is the phase-space density. When considering an equation about $N_{e^\pm}$ it yields a Fokker-Planck equation with the two coefficients:
\begin{eqnarray} 
A_{e^\pm}^{\rm{acc}} &=& -\frac{1}{p\gamma} \partial_\gamma \left( \frac{p^3}{\gamma} D^{\rm{diff}} \right)~, \\
D_{e^\pm}^{\rm{acc}} &=& 2 \frac{p^2}{\gamma^2} D^{\rm{diff}} ~.
\end{eqnarray}
We assume a Fermi-like process for particles with an energy above some minimal energy, and we use $D^{\rm{diff}} = p^2e^{-(p_c/p)^a}/2t_{\rm{acc}}$, where $p_c$ is the threshold momentum, $a$ is the threshold width (we use typically $a=3$), and $t_{\rm{acc}}$ is the typical acceleration time of the particles \citep{Katar06}. The FP coefficients are then:
\begin{eqnarray} 
A_{e^\pm}^{\rm{acc}} &=& -\frac{1}{2t_{\rm{acc}}} \frac{p^2}{\gamma^3}\left(5+4p^2 +a\gamma^2 \left(\frac{p_c}{p}\right)^a \right)e^{-(p_c/p)^a} ~,\\
D_{e^\pm}^{\rm{acc}} &=& \frac{1}{t_{\rm{acc}}} \frac{p^4}{\gamma^2} e^{-(p_c/p)^a} ~.
\end{eqnarray}
The precise values of acceleration time and the threshold frequency depend on the microphysics and turbulent properties of the plasma, which are poorly known. We set instead the total energy injected into accelerated particles by defining a constant compactness parameter:
\begin{equation}
l_{\rm{acc}} = \frac{4\pi}{3} \frac{R^2 \sigma_T}{c} \int -A_{e^\pm}^{\rm{acc}} N_e dp
\end{equation}
and compute the corresponding acceleration time at each time step.

\subsection{Bremsstrahlung emission}
The bremsstrahlung process has several contributions to the system evolution: it produces additional soft photons that can then be up-scattered by high energy particles, it cools down emitting, high energy particles and, in the absorbed part, it heats low energy particles. In arbitrary plasmas, there are three different contributions: lepton-proton ($e-p$), electron-electron or positron-positron ($e-e$), and electron-positron ($e^--e^+$) bremsstrahlung.

Electron-proton self-absorbed bremsstrahlung is included in the code. Proton are assumed to have non-relativistic temperature and to be at rest in the plasma frame and the emission is generated by the motion of leptons in the external electrostatic potential of protons. The situation is formally the same as in the case of synchrotron emission, which is generated by the motion of leptons in an external magnetic field, and a similar formalism can be used \citep{LeRoux60,GS91}. The exact interaction cross section $\sigma^0_{\rm{ep}}$ valid for all lepton energies \citep{Heitler54,JR76} is used to compute the emissivity spectrum $j_{\rm{ep}}$ (erg $s^{-1}$ Hz$^{-1}$) of individual electrons and the absorption cross section $\sigma_{\rm{ep}}$ (cm$^2$):
\begin{eqnarray}
j_{\rm{ep}}(p,\nu) &=& h\nu \beta c\sigma_{\rm{ep}}^0(p,\nu) N_p^{tot}\\
\sigma_{\rm{ep}}(p,\nu) &=&  \frac{1}{4\pi} \frac{1}{2m_e\nu^2} \frac{1}{p^2} \partial_p \left[ p\gamma j_{\rm{ep}}(p,\nu) \right] 
\end{eqnarray}
where $N_p^{tot}$ is the proton density, $\beta=v/c$ is the lepton velocity, and $\sigma_{\rm{ep}}$ is evaluated numerically. Approximations of the total loss rate and the spectrum emitted by a thermal plasma are recovered by integrating $j_{\rm{ep}}$ over a thermal distribution of leptons. Then, the evolution equations for particles and photons are derived exactly as in the case of synchrotron emission (Eqs. \ref{eq_Inu}-\ref{eq_sync3}).

Electron-proton bremsstrahlung is the dominant contribution in low energy, e-p plasmas. In low-energy, pair plasmas the $e^+-e^-$ process dominates, whereas at high temperatures the major contribution originates in $e-e$ bremsstrahlung. Differential cross sections for $e-e$ and $e^--e^+$ bremsstrahlung can be found in various regimes, either in the rest frame of one of the leptons or of the centre of mass of the two interacting particles \citep{Heitler54,Alexanian68,Haug75,Haug85}. However, there is no formula in the frame of the plasma for the cross section integrated over all directions of the emitted photon; simpler thermal approximations are also often irrelevant since the cross section typically increases with particle energy and non thermal emission of high energy particles often dominates the overall bremsstrahlung emission. In principle, these difficulties could be overcome numerically.

However, in many astrophysical cases bremsstrahlung emission is insignificant. For plasmas in thermal equilibrium, simple approximations were proposed for the cooling rates, which allowed for comparison with other processes \citep[e.g.][]{Gould80,Stepney83b}. It was found that bremsstrahlung cooling dominates over pair-annihilation cooling and Coulomb relaxation only of highly relativistic temperatures: $k_BT\gtrsim1$ MeV \citep[respectively]{Svensson82b,Stepney83a}. By integrating the synchrotron and Compton cooling rates over a hot thermal distribution of particles ($k_BT$ = 1 MeV) of radiation energy density $U_\nu$, we find that, for plasmas with optical depth of the order of unity, bremsstrahlung emission dominates only when $R\sigma_TU_\nu/m_ec^2+l_B \lesssim 4\times10^{-3}$, that is for unmagnetized, photon-starved plasmas. For these compactness parameters, only hotter plasmas have a significant bremsstrahlung contribution, although, for astrophysical sources, these high temperatures are unrealistic since pair production and annihilation tend to prevent temperatures reaching above a few hundreds keV \citep{Svensson84}. Similarly, it is has been shown that non-thermal particles emit primarily synchrotron radiation, even for weak magnetic fields \citep{Wardzinski00,Coppi92}.

For these reasons, $e-e$ and $e^--e^+$ bremsstrahlung have not been included in our code and modelling of unmagnetized, highly relativistic plasmas with a weak radiation field is postponed to future work.


\section{Numerical methods}\label{sec:numsol}
Including all processes outlined above, the total physical system is described by the following set of 3 integro-differential equations:
\begin{eqnarray}
\partial_t N_\nu &=&  S_\nu  - P_\nu N_\nu + \partial_\omega \left[ A_\nu N_\nu\right] + \frac{1}{2} \partial^2_{\omega^2} \left[ D_\nu N_\nu \right] ~, \label{the_eq1} \\
\partial_t N_{e^\pm} &=& S_{e^\pm} - P_{e^\pm} N_{e^\pm} + \partial_p \left[ \gamma/p A_{e^\pm} N_{e^\pm} \right]  \nonumber \\
&& \quad\quad\quad\quad\quad\quad + \frac{1}{2} \partial_p \left[ \frac{\gamma}{p} \partial_p \left( \frac{\gamma}{p} D_{e^\pm} N_{e^\pm} \right) \right] ~. \label{the_eq2}
\end{eqnarray}
The source terms $S_{\nu}(t,\nu,N_\nu,N_{e^-},N_{e^+})$ and $S_{e^\pm}(t,p,N_\nu,N_{e^-},N_{e^+})$ combine the contributions of injection, Compton scattering (treated in the integral approach), annihilation/production, bremsstrahlung, and synchrotron emission:
\begin{eqnarray}
S_\nu &=& \dot{N}^{\rm{inj}}_\nu(t,\nu) \nonumber  \\
&+& \int_{0}^{\infty} d\nu_0 \int_{p_c(\nu_0)}^{\infty} dp_0 N_e(p_0) N_\nu(\nu_0) c \sigma^{\rm{c}}(p_0,\nu_0;\nu)   \nonumber \\
&+& \int \hspace{-.3cm}\int_{0}^{\infty}  N_{e^-}(p_-) N_{e^+}(p_+)c \sigma_{\rm{a}}(p_-,p_+; \nu ) dp_- dp_+ \nonumber \\
&+& \frac{1}{h\nu} \int N_e (j_s+j_{\rm{ep}}) dp ~, \\
S_{e^\pm} &=& \dot{N}^{\rm{inj}}_{e^\pm}(t,p) \nonumber  \\
&+& \int_{0}^{\infty} dp_0 \int_{\nu_c(p_0)}^{\infty} d\nu_0 N_e(p_0) N_\nu(\nu_0) c \sigma^{\rm{c}}(p_0,\nu_0;\nu(p))   \nonumber \\
&+& \int \hspace{-.3cm}\int_{0}^{\infty}  N_{\nu}(\nu_1) N_{\nu}(\nu_2)c \sigma_{\rm{p}}(\nu_1,\nu_2; p ) d\nu_1 d\nu_2 ~.
\end{eqnarray} 
The loss terms  $P_{\nu}(t,\nu,N_\nu,N_{e^-},N_{e^+})$ and $P_{e^\pm}(t,p,N_\nu,N_{e^\mp})$ also combine contributions from escape, Compton scattering, pair production/annihilation, Bremstrahlung, and synchrotron absorption:
\begin{eqnarray}
P_\nu &=& p^{\rm{esc}}_\nu(\nu,N_{e^-},N_{e^+}) + \int_{p_c(\nu)}^{\infty}  N_e(p) c \sigma^{\rm{c}}_0(p,\nu) dp  \nonumber \\
&+& \int N_{\nu}(\nu') c \sigma_0^{\rm{p}}(\nu,\nu') d\nu'  +  \int N_e c (\sigma_s+\sigma_{\rm{ep}}) dp ~,\\
P_{e^\pm} &=& p^{\rm{esc}}_{e^\pm}(p) + \int_{\nu_c(p_0)}^{\infty} N_\nu(\nu) c \sigma^{\rm{c}}_0(p,\nu) d\nu  \nonumber \\
&+& \int  N_{e^\mp}(p')c \sigma_0^{\rm{a}}(p,p') dp'  ~.
\end{eqnarray} 
\\
The total Fokker-Planck coefficients are then the sums of the individual coefficients defined for each process:
\begin{eqnarray}
A_\nu &=& -\int_{0}^{p_c(\nu)} N_e(p) c\sigma_1^c(\nu,p) dp ~,  \\
D_\nu &=& \int_{0}^{p_c(\nu)} N_e(p) c\sigma_2^c(\nu,p) dp 
\end{eqnarray}
and 
\begin{eqnarray}
A_{e^\pm} &=& \frac{1}{m_ec^2} \int \left( (j_s+j_{\rm{ep}}) -c(\sigma_s+\sigma_{\rm{ep}}) h\nu N_\nu \right) d\nu  \nonumber \\
&+&   \int_{0}^{\nu_c(p)} \hspace{-.5cm}N_\nu(\nu) c\sigma_1^c(\nu,p) d\nu \nonumber \\ 
&+&  \int N_{e^\mp}(p') a_e(p,p') dp' +  \int N_{p}(p') a_p(p,p') dp' \nonumber \\
&+& A_{e^\pm}^{\rm{e-p}} + A_{e^\pm}^{\rm{acc}} ~, \\
D_{e^\pm} &=& \frac{1}{4\pi}\frac{h}{m_e^2c}\int \frac{(j_s+j_{\rm{ep}}) N_\nu }{\nu} d\nu  \nonumber \\
&+& \int_0^{\nu_c(p)} N_\nu(\nu) c\sigma_2^c(\nu,p) d\nu \nonumber \\
&+&  \int N_{e^\mp}(p') d_e(p,p') dp' +  \int N_{p}(p') d_p(p,p') dp' \nonumber \\
&+& D_{e^\pm}^{\rm{e-p}} + D_{e^\pm}^{\rm{acc}} ~.
\end{eqnarray}
In this section, we describe the numerical strategy used to solve these equations.



\subsection{Tables}
Solving Eqs. \ref{the_eq1} and \ref{the_eq2} for the aforementioned physical processes involves significant application of the many cross sections ($j_s,\sigma_s, j_{\rm{ep}},\sigma_{\rm{ep}},\sigma_{\rm{c}},\sigma^{\rm{c}}_0,\sigma^{\rm{c}}_1,\sigma^{\rm{c}}_2,\sigma_{\rm{a}},\sigma^{\rm{a}}_0,\sigma_{\rm{p}},\sigma^{\rm{p}}_0,a_e,d_e,a_p,d_p$). The Compton differential cross section $\sigma^{\rm{c}}(\nu_0,p_0\rightarrow\nu)$ is a three-entry table that contains typically over tens of millions of elements, as is also the case for the differential cross sections for pair production/annihilation. These coefficients are evaluated once at the beginning of the simulation and stored into tables. This enables faster computation, although the memory requirements can become significant as the grid resolution increases. For instance, resolutions of above 256 points per grid require more than 100 Mo of RAM to store only one of these tables, which can be a limiting factor for some desktop computers.

\subsection{Boundaries}
The total set of equations to be solved is given by an integro-differential system. To account for particles and photons that have energies beyond the limits of the grids (very low- or very high energy particles/photons), specific conditions must be set at the grid boundaries $\omega_{\rm{min}}$, $\omega_{\rm{max}}$, $p_{\rm{min}}$, and  $p_{\rm{max}}$. For the differential Fokker-Planck part of the equation which corresponds to a  {\it local} operator, the boundary conditions set values only for the {\it ghost} bins just behind the boundaries and are used to define the derivative at the boundaries. For the integral part, they include the physics of all particles and photons outside the grids. We have chosen to use wall-type boundary conditions. These conditions do not allow particles/photons to travel in and out of the grids and conserve their total number precisely. In the Fokker-Planck part, it corresponds to a zero-flux condition. For the integral part, a specific derivation of the differential cross section is completed: the total probability that particles/photons are scattered or produced outside the grids is summed and added to the probability that they are scattered or produced at the final bin of the relevant boundary. As a result, all particles/photons remain inside the grids. These conditions are of course artificial and can generate spurious effects at the boundaries. While most particles/photons have energies inside the grids, these effects will, however, remain small.

For example, although the total number of particles/photons is conserved, the no flux condition for the FP part of the equations introduces a small energy flux at the boundaries which can be evaluated easily, that is the energy losses are $dE_\omega/dt = m_ec^2 \left[ D_{\omega} N_{\omega} \right]_{\omega_{\rm{min}}}^{\omega_{\rm{max}}}$ and $dE_{e^\pm}/dt = m_ec^2 \left[ \gamma/p D_{e^\pm} N_{e^\pm} \right]_{p_{\rm{min}}}^{p_{\rm{max}}}$ for photons and leptons respectively. When the grid is not sufficiently large, this effect can introduce significant errors. By selecting the boundaries far from the bulk of the distribution however, the distributions $N_\omega$ and $N_{e^\pm}$ vanish and the losses are negligible. 

\subsection{Numerical solver}
Solving the time-dependent problem is challenging for a number of reasons. 

First, the problem involves many different scales of energy and time spanning many orders of magnitude. This implies subtracting very large numbers or multiplying very small numbers by very large ones, which can lead to numerical accuracy issues. 

This also involves considering very short timescales. For instance, when low energy photons are scattered by high energy particles, they absorb a significant amount of energy instantaneously, which must be modelled with very small time steps. If the problem was linear and differential, the maximal time step required to guarantee the convergence and stability of an explicit scheme would be set by the Courant condition \citep{CFL}. In that particular case, our simple scheme for the photon equation would be stable if $\delta t < \rm{min}\left\{\delta \omega /  A_\nu ; 2 (\delta \omega)^2/D_\nu \right\}$, where $\delta \omega $ is the bin size and the minimum is computed over the entire grid. Similarly, a condition depending on the momentum bins would be set for the stability of the equation for particles. When logarithmic grids are used, the time step is set to a small value by the small bins at the low energy part of the grids. The equation is far more complicated, and there is no mathematical justification for using the Courant condition. However, the main idea remains that when the grids decline to low energy, the time step required to make an explicit scheme stable, quickly becomes too small to follow the evolution on the dynamical timescale $R/c$.

In these cases, implicit schemes would be more efficient, since they are always stable. Implicit schemes are easy to implement and efficient only for local, linear problems. When the problem is linear, the solver only has to inverse a matrix. For local problems such as differential ones, the matrix is sparse and rapidly inverted. However, the problem is highly non-local. For example, the integral approach of Compton scattering describes events in which photons in some energy bin can be scattered to some distant bin by a single interaction. As a result, the evolution in the photon distribution at some energy is governed not only by neighbouring bins, but by the entire grid. The corresponding matrix is dense and its inversion becomes more time consuming. In addition, the problem is highly non-linear, so that there is no {\it exact} implicit solution to such a problem. 

Considering these previous remarks, we implemented a semi-implicit scheme. By keeping the other distributions fixed, we solved the equation for the distribution $N$ of one species (photons, electrons, or positrons) in the following way:
\begin{eqnarray}
\frac{N^{n+1}-N^n}{\delta t} &=&  S^n  - P^n N^{n+1} \nonumber \\
  &&\quad + \partial_x \left[ A^n N^{n+1} \right]  + \frac{1}{2} \partial^2_{x^2} \left[ D^n N^{n+1} \right]  \label{eq_n}
\end{eqnarray}
where the exponents $n$ and $n+1$ indicate the distributions and coefficients at two consecutive time steps. The scheme is not fully implicit but the equations for $N^{n+1}_\nu$, $N^{n+1}_{e^-}$ and $N^{n+1}_{e^+}$ can be solved easily by inverting a simple multi-diagonal matrix. This scheme only conserves the number of particles/photons and their energy to truncations errors. As explained earlier, using grids spanning many orders of magnitude introduces accuracy issues, which become severe when number and energy are only conserved to truncation errors. For that reason, the code iterates the 3 equations represented by Eq. \ref{eq_n} at each time step until the conservation laws are satisfied to some specific precision. 

Two spatial schemes were implemented to solve Eq. \ref{eq_n}: the upwind Chang-Cooper method \citep{CC70} and a more straightforward method that we developed. The former is only accurate to first order in space, but was shown to contain the most superior numerical properties for solving Fokker-Planck equations with a few choices of {\it constant} coefficients \citep{PP96}. Compared with higher order schemes, it is more diffusive, i.e. more stable but less accurate. When radiative processes are involved, accuracy is important, and therefore we decided to use a second-order accurate scheme to estimate the derivatives\footnote{The scheme is accurate to second order only when the grid is linear. Using a logarithmic grid as we do here actually reduces the accuracy. Nevertheless, numerical experiments with the code have shown that this scheme is more accurate than the Chang-Cooper method.}.  This scheme is based on the use of two energy grids for each distribution (namely the centres and the faces of the bins) and the derivatives are computed as follows: $\left(\partial_x f\right)_i=(f_{i+1/2}-f_{i-1/2})/dx_i$ and $\left(\partial^2_x f \right)_i= ( (f_{i+1}-f_{i})/dx_{i+1/2} - (f_{i}-f_{i-1})/dx_{i-1/2})/dx_i$. All quantities at the bin boundaries are computed by linear interpolation: $f_{i+1/2}=(f_i+f_{i+1})/2$, except the $(p/\gamma)_{i+1/2}$ factor in Eq. \ref{eq_FP} for particles, which is computed separately as $(p_i+p_{i+1})/(\gamma_i+\gamma_{i+1})$ to ensure accurate energy conservation. This simple centred scheme conserves the total number of particles and photons, and the total energy more precisely, although is less stable. All examples presented were performed using the latter scheme.\\

\subsection{Computation time}
Tests and applications shown in this work were completed with medium energy resolution for which $n_\nu = 128-256$ and $n_p=256$, i.e. more than 20 bins per decade in particle momentum and more than 10 bins per decade in photon frequency. Compared with former codes \citep{LZ87, Coppi92} that assume low energy particles are purely thermal, our code typically solves the equations for twice as many bins of particle momentum. In addition, as presented, it adopts very few approximations and only when they are valid. The code is therefore slower than some previous codes. All results presented were derived with desktop computers with 1GHz processors and 512 MB RAM. Calculations have duration typically of between a few seconds and one hour, the most time-consuming step being the computation of the multiple integrals of the Compton and pair production/annihilation cross sections over the distributions. 


\section{Tests and applications}\label{sec:testsandapps}
We present a first few applications of the code to check its capabilities and illustrate the problems that can be addressed. Most examples presented consist of comparisons with previous work, although we also address a few more issues, such as the threshold for particle acceleration. 
\subsection{Model with external soft photons}
\label{coppi_sec}
\begin{figure}[h!]
\begin{center}
\includegraphics[scale=.5]{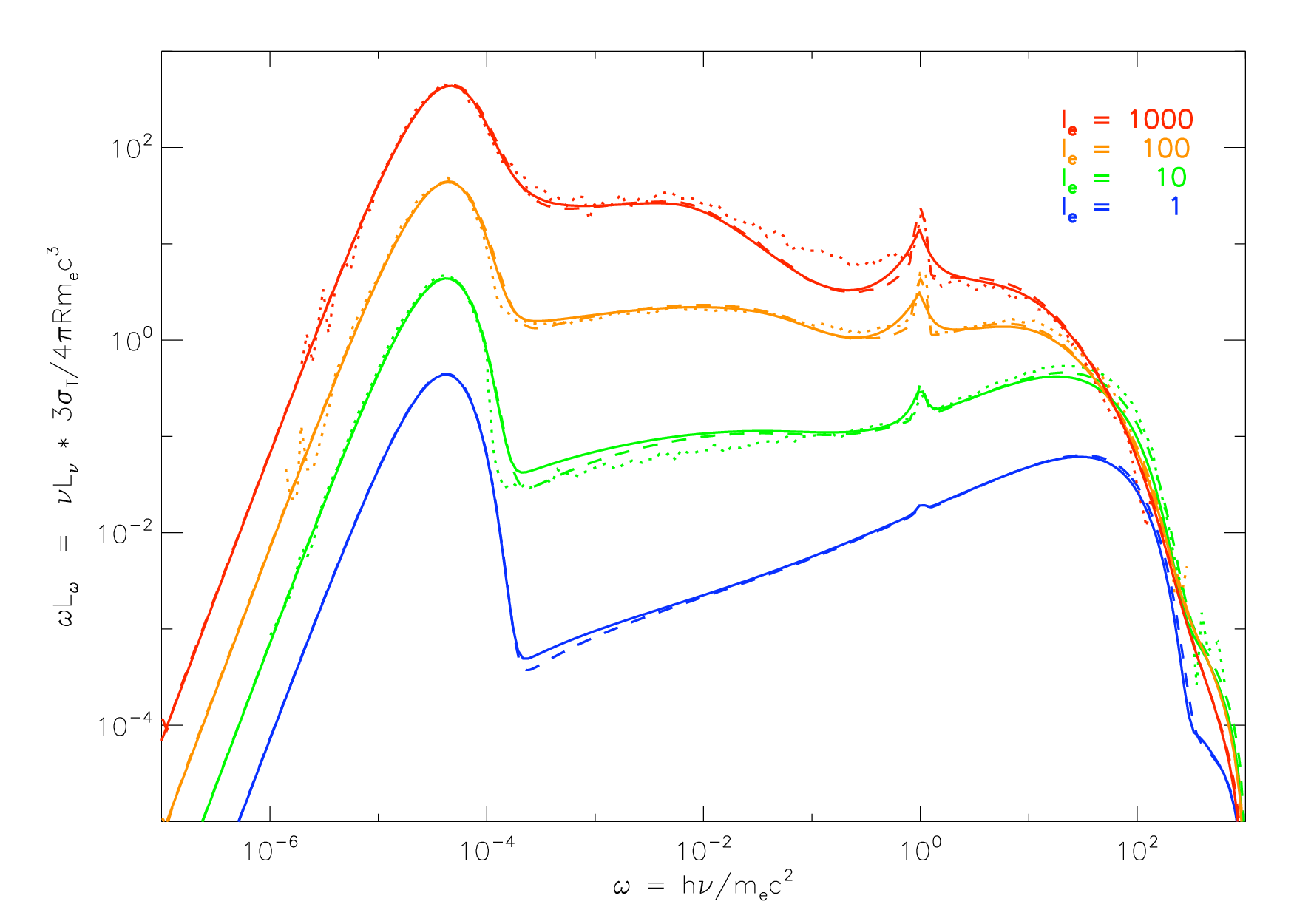}
\caption{Photon spectra for models with external soft photons ($l_{e}=l_{e^-}+l_{e^+}=1,10,100,1000$ from lower to higher curve and $l_\nu=2.5 l_{e}$). Solid lines: this work; dashed lines: results with the EQPAIR code; and dotted lines: results of Monte Carlo simulation for spherical geometry \citep{Stern95}. The spectra from EQPAIR were normalized to match the other ones. To simplify the comparison, the escape probability $p_\nu^{\rm{esc}}=1/(1+\tau f)$ of \citet{Coppi92} is used in this figure.} \label{Coppi92a}
\end{center}
\end{figure}

Significant effort has already been expended on studying steady state solutions of unmagnetized sources. We only reiterate some known results in checking the code computational capacities. As a first case, we reproduced the results of Fig. 1 in \citet{Coppi92} with parameters typical of AGN. This case was modelled by injecting mono-energetic $e^{+}-e^{-}$ pairs at $\gamma=10^3$ and soft photons as a black body of temperature $k_BT=1.07\times 10^{-5} m_ec^2$. In this unmagnetized source, the length scale is unimportant. All effects were included except additional particle acceleration and e-p Coulomb scattering and e-p bremsstrahlung radiation. Figure (\ref{Coppi92a}) shows spectra obtained with the code and comparisons with previous work. 
Examples of output are also presented in Table \ref{Coppi_tab1}. The temperature presented in this table was computed using Eq. (2.8) of \citet{Coppi92} only for the thermal part of the particle distribution.
\begin{table}[h!]
\begin{center}
\begin{tabular}{c|c|ccccc|}
$l_e$ & work & PY$_3$ & $\tau_e$ & $\theta_3$ & $l_X/l_e$ & $ \alpha_{2-10}$ \\  \hline 
& Coppi92 & 1.7 & 0.047 &  7.9 & --- & 0.637\\
$1$ & EQPAIR & --- & 0.078 & 12.6  & 0.0582  &  0.614 \\
& This work & 1.70 & 0.059 & 8.64  & 0.0611 & 0.633 \\ \hline
& Coppi92 & 23 & 0.502 & 5.7  & --- & 0.863\\
$10$ & EQPAIR &  --- & 0.508 & 8.27  &2.70 & 0.854  \\
& This work & 22.6 & 0.548 & 4.50  & 3.00 & 0.901 \\ \hline
 & Coppi92 & 87 & 3.34 & 2.2  & ---  & 0.979 \\
$10^2$ & EQPAIR &  ---& 3.17 &  2.56 & 63.9 & 0.996 \\
& This work & 80 & 3.22 & 1.96  & 61.9 & 1.02 \\ \hline
& Coppi92 &  120 & 12.4 & 0.62  & --- & 1.42 \\
$10^3$ & EQPAIR & ---  & 11.95 & 0.567  & 644 & 1.39  \\
& This work & 112 & 12.01 & 0.44  & 628 & 1.41 \\ 
 \hline
\end{tabular}
\caption{Models with external soft photons: comparison with previous results. PY$_3$ is $10^3$ times the pair yield defined in \citet{Coppi92}. $\tau_e=R \sigma_T\int (N_{e^-}+N_{e^+})dp$ is the total Thomson optical depth. $\theta_3 = 10^3 \times k_B T/m_ec^2$ is the temperature of the thermal part of the distribution. $l_X/l_e$ is the ratio of the X-ray luminosity in the 2-10 keV band to the injection compactness parameter. $\alpha_{2-10}$ is the spectral index in the same energy band.} \label{Coppi_tab1}
\end{center}
\end{table}
The results are fully consistent with those computed with the latest version of the public EQPAIR code. The major deviations appear at the annihilation line for large optical depths and luminosities ($l_e=100$ and $1000$). 

\begin{figure}[h!]
\begin{center}
\includegraphics[scale=.5]{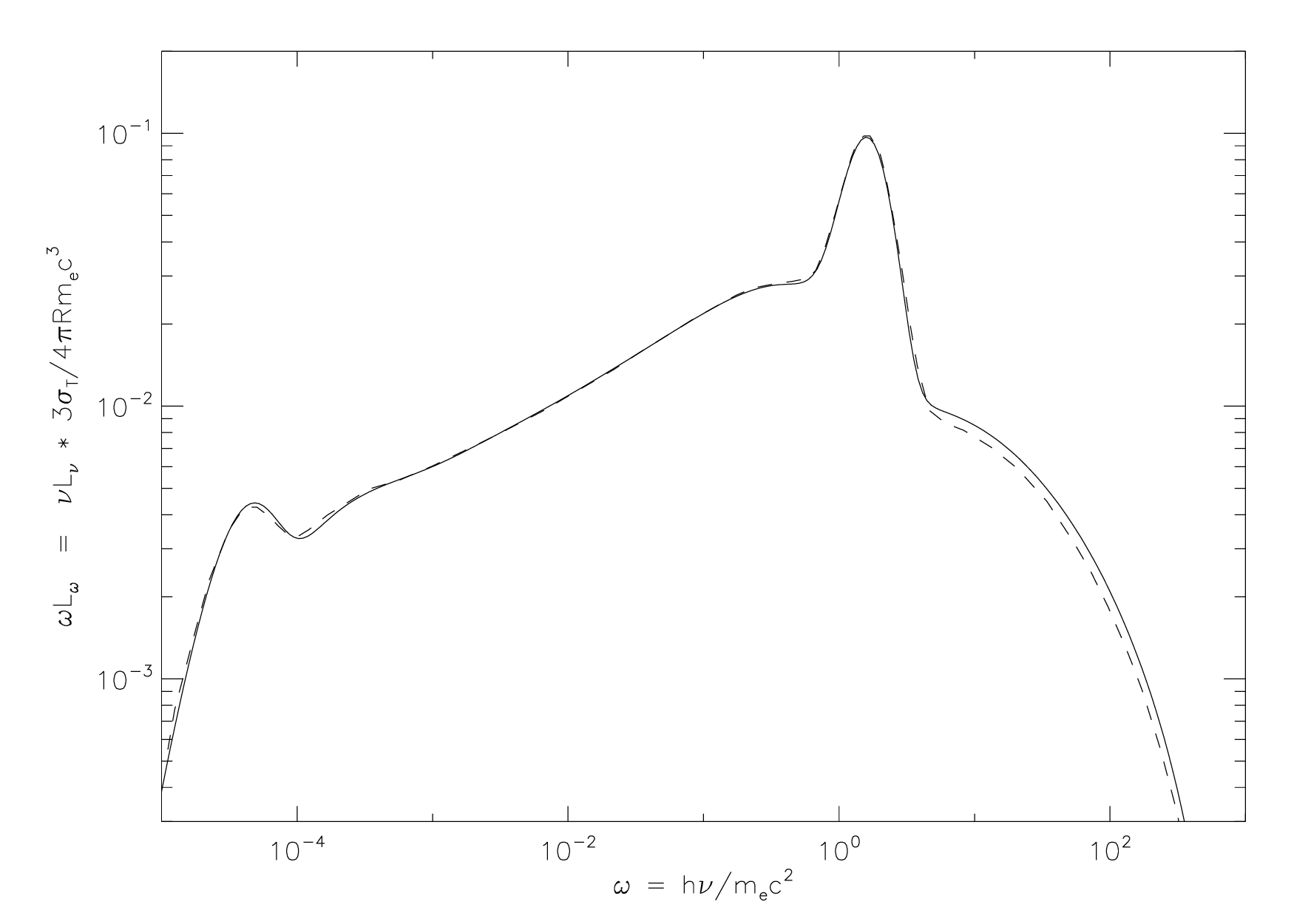}
\includegraphics[scale=.5]{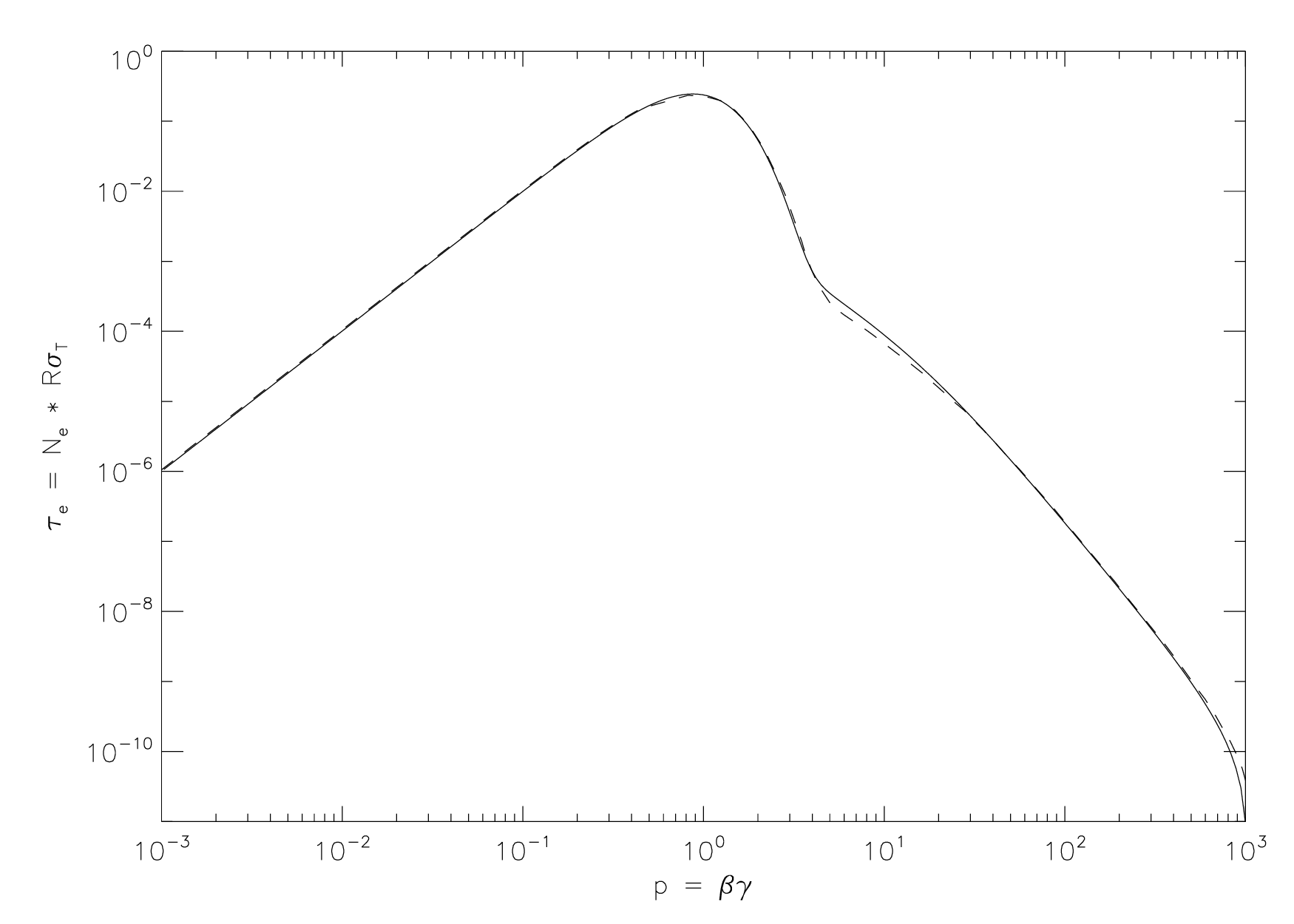}
\caption{Photons spectrum (upper panel) and particle distribution (lower panel) for power-law particle injection (with $l_e=1$, $\gamma_{\rm{min}}=1.4$, $\gamma_{\rm{max}}=10^3$, $\Gamma=2.4$) and for $l_\nu=0.03$. The results with the code are shown in solid line and the results of \citet{Coppi92} are in dashed lines. Since the particle distribution at low energy in \citet{Coppi92} is assumed to be thermal but not resolved, we extended it down to the grid boundary with a Maxwell distribution for comparison.} \label{Coppi92fig2}
\end{center}
\end{figure}

When $\tau_e/(l_e+l_\nu) \gtrsim 1$, effects of pair annihilation and Coulomb scattering become significant. We investigated this regime by reproducing the results presented in Fig. 2 of \citet{Coppi92}. This simulation had the same input parameters as previously, except that particles were injected with a power law distribution ($\gamma_{\rm{min}}=1.4$, $\gamma_{\rm{max}}=10^3$, $\Gamma=2.4$, $l_e=1$) and $l_\nu=0.03$. The particle distribution and photon spectrum for such a case are plotted on Fig. \ref{Coppi92fig2}.
The corresponding output parameters are listed in Table \ref{Coppi92_outputs}.
\begin{table}[h!]
\begin{center}
\begin{tabular}{l|ccccc|}
& PY$_3$ & $\tau_e$ & $\theta_3$ & $l_X/l_e$ & $ \alpha_{2-10}$ \\  \hline 
Coppi92 & 3.54  & 0.644 &  306 & $0.077^*$ & 0.726 \\
 This work & 3.27 &  0.624 &  273 & 0.0733 &0.717 \\  \hline
\end{tabular}
\caption{Output parameters for $l_e=1$ and $l_\nu=0.03$. Same as table \ref{Coppi_tab1}. *Note that the luminosity obtained by \citet{Coppi92} was multiplied by 10 to correct what we think is a typo.} \label{Coppi92_outputs}
\end{center}
\end{table}
Again, good agreement is achieved, particularly for the particle distribution, confirming that, in this peculiar case, the approximations made by \citet{Coppi92} were valid. The only difference occurs at high energy in the photon spectrum. 

\subsection{e-p Coulomb like heating}
\begin{figure}[h!]
\begin{center}
\includegraphics[scale=.5]{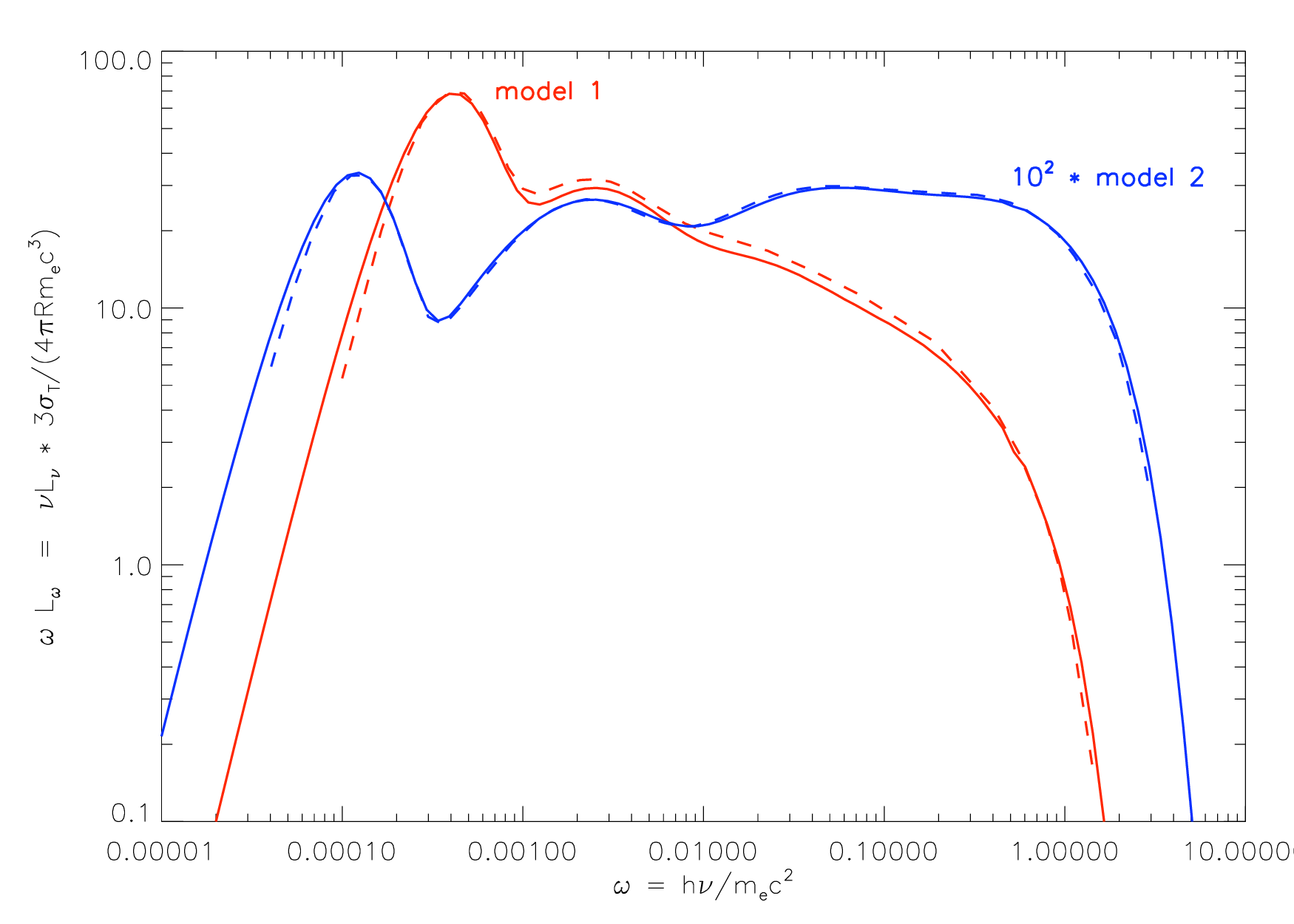}
\includegraphics[scale=.5]{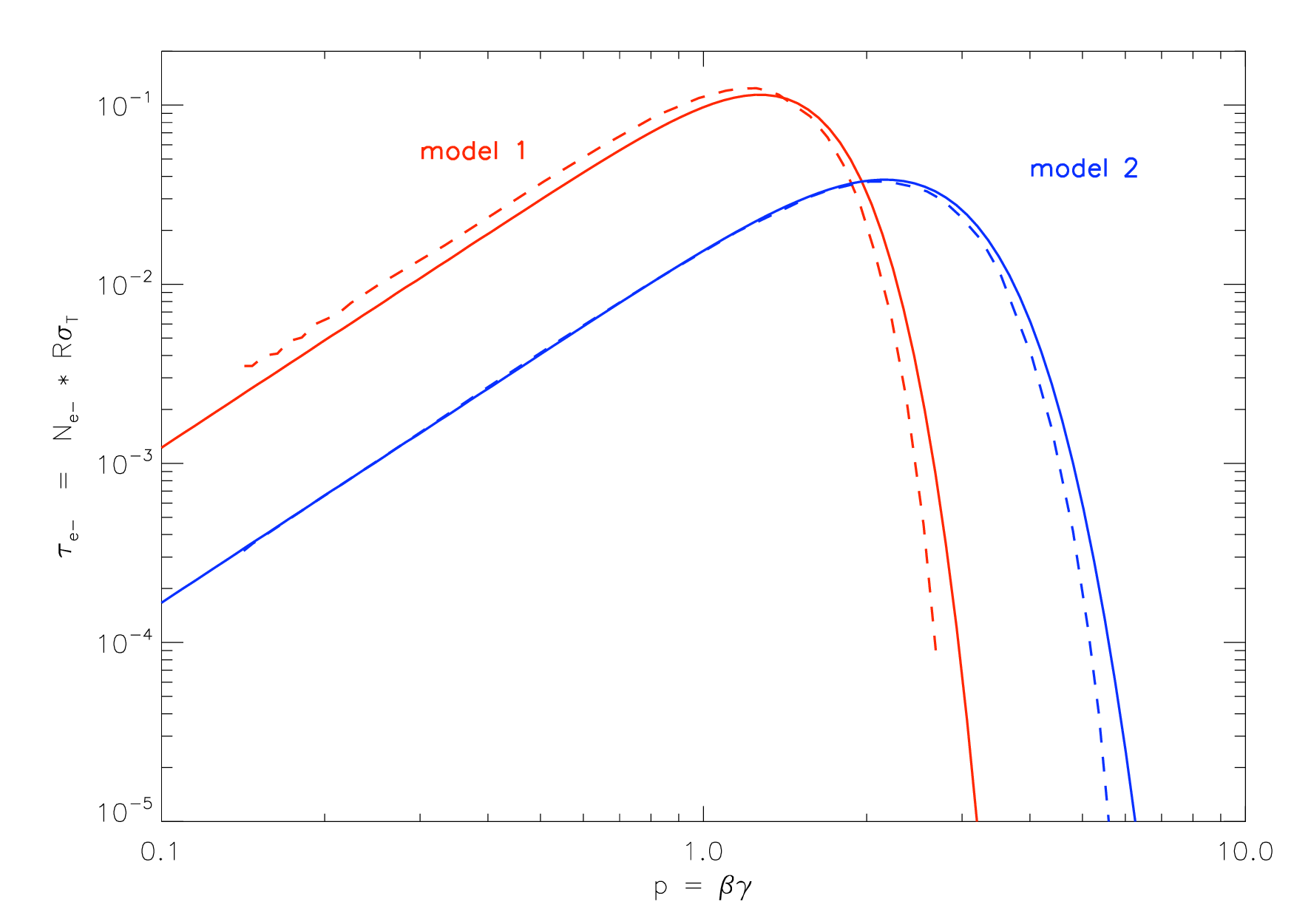}
\caption{steady state spectra (upper panel) and electron distribution (lower panel) for e-p Coulomb-like heating. Solid and dashed lines show the results of this work and \citet{NM98}, respectively. Input and output parameters are listed in Tables \ref{nm98_inputs} and \ref{nm98_outputs}, respectively.} \label{NM98}
\end{center}
\end{figure}

We investigated the effect of coulomb-type heating and compared the results with those of \citet{NM98}.  We consider an unmagnetized source heated by a Coulomb-like process. We assumed a closed system (no injection nor loss of particles) with a black-body soft-photon injection and studied its evolution under the effects of Compton scattering, e-e Coulomb exchange, pair production/annihilation, and e-p Coulomb-like heating. The proton temperature characterising the final process was set to be 20 MeV. Two different cases were considered, the parameters of which are given in Table \ref{NM98}.
\begin{table}[h!]
\begin{center}
\begin{tabular}{c|cccc|}
 & $l_c$ & $l_\nu$ & $\tau_{e^-}^0$ & $\Theta_\nu $ \\ \hline
 Model 1 & 420 & 420 & 0.05 & $10^{-4}$ \\
 Model 2 & 8.4 & 2.1& 0.02 & $3\times10^{-5}$  \\
 \hline
\end{tabular}
\caption{Input parameters for the runs on the study of the e-p Coulomb like heating. $\tau_{e^-}^0$ is the initial electron Thomson optical depth. $\Theta_\nu=k_BT/m_ec^2$ is the temperature of the black body used for soft photon injection.} \label{nm98_inputs}
\end{center}
\end{table}
Regardless of the initial particle distribution, it always evolved to the same steady solution. The thermalisation time however depended upon the precise initial shape. Distributions in the transient phase were presented by \citet{NM98} and we derived similar results.
Figure \ref{NM98} shows the steady distributions and spectra for both cases, and compares the result with previous work. 
The steady states correspond to quasi-thermal distributions. Both the e-e Coulomb exchange and the e-p Coulomb-like interactions thermalize the plasma efficiently. In addition, the e-p Coulomb-like interactions heat the particle distribution. 

Our steady spectra are fully consistent with previous results: the Compton orders have the same amplitude, and the high energy spectrum breaks at the same energy. Some output parameters are also given in Table \ref{nm98_outputs}. Our results confirm qualitatively the results by \citet{NM98}. The steady distributions have properties similar to those of a Maxwell-Boltzmann distribution but are narrower. Our results however correspond to systematically hotter distributions and a larger optical depth in the most energetic case (model 2).
\begin{table}[h!]
\begin{center}
\begin{tabular}{c|l|cccc|}
Model & work & $\tau_{e^-}$ & $\tau_{e^+}$ & $\left<E_c\right>$ & $\eta_3$ \\  \hline 
1 & NM98 & 0.135 & 0.085 & 0.58 &  $2.4 $ \\
 &This work & 0.132 & 0.082 & 0.65 & $110$ \\ \hline 
2& NM98 & 0.080 & 0.060 & 1.46 & 0.44 \\
& This work & 0.086 & 0.066 & 1.53 & 17   \\
 \hline
\end{tabular}
\caption{Output parameters. $\tau_{e^\pm}$ are the electron and positron Thomson optical depths.  $\left<E_c\right> = \int (\gamma-1) N_{e^-} dp /\int N_{e^-} dp $ is the averaged kinetic energy of the electron distribution. $\eta = \eta_3\times 10^{3} $ is the heating efficiency defined by Eq. \ref{lh}.} \label{nm98_outputs}
\end{center}
\end{table}

The heating efficiency coefficient is quite large: $\eta \approx 10^4-10^5$. This emphasizes the inefficiency of the e-p Coulomb collisions for example in cases typical of Seyfert galaxies. When the Thomson optical depth is far smaller than unity, they should be several orders of magnitude more efficient to reach the required heating rate. The efficiency coefficients found in this work are one or two orders of magnitude higher than those derived by \citet{NM98}. Given the good agreement in the shape and normalisation of the distributions and escaping spectra, we believe that the different efficiencies are the result of a typo in their paper.  We further checked our e-p Coulomb heating rate against the results of \citet{DML96}, and also against simple analytic approximations, and found excellent agreement. 

\subsection{Models with synchrotron soft photons}
\begin{figure}[h!]
\begin{center}
\includegraphics[scale=.5]{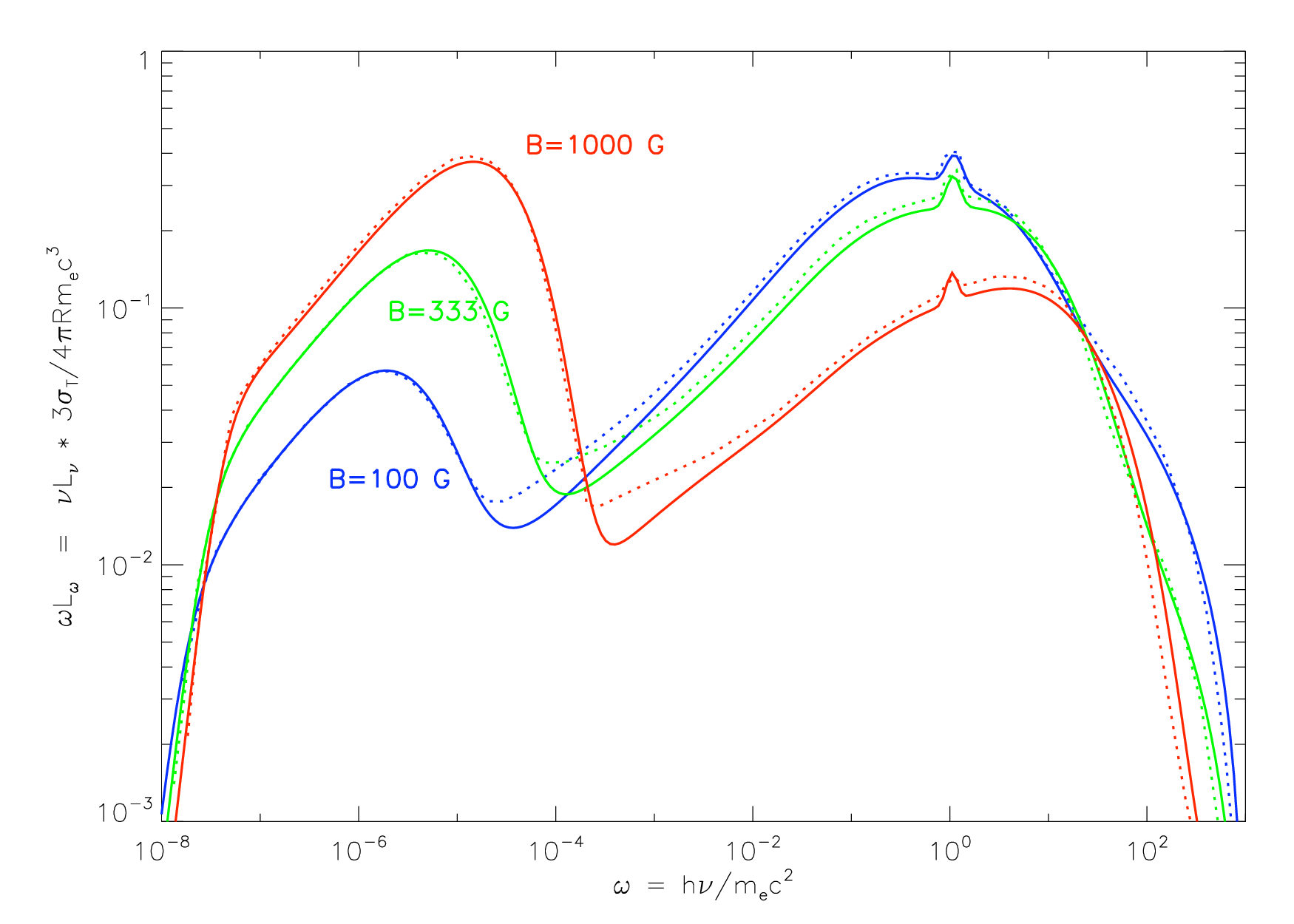}
\caption{Synchrotron self-compton spectra. Spectra are shown for 3 magnetic field strengths ($l_B=3.2 \times 10^{-2},3.2 \times 10^{-1},3.2$), from results of this work (solid lines) and \citet{Coppi92} (dotted lines). Here, $l_e=10$ and $R=10^{14}$ cm. } \label{Coppi92fig4}
\end{center}
\end{figure}

Magnetized models have an additional source of soft photons, which is the synchrotron emission from high energy particles. We investigated this case by studying generic cases of magnetised sources with no external source of soft photons and compared our results with those presented in Fig. 4 of \citet{Coppi92}. Particles were injected at high energy ($\gamma=10^3$). We assumed they were trapped by the magnetic field and did not escape. The equilibrium was therefore balanced by pair annihilation. A few synchrotron self-compton spectra are shown in Fig. \ref{Coppi92fig4} and compared with previous work. Other output parameters are listed in Table \ref{Coppi92_outputs2}.
\begin{table}[h!]
\begin{center}
\begin{tabular}{c|c|ccccc|}
B &work  & PY$_3$ & $\tau_e$ & $\theta_3$ & $l_X/l_e$ & $ \alpha_{2-10}$ \\  \hline 
100 & Coppi92 &  21.0 & 0.508  & 37.3 & 82.4 & 0.578 \\
            & This work &  18.9 &  0.513 & 64.1 & 72.6 & 0.561 \\ \hline
300 & Coppi92 & $15.4^*$  & 0.422 & $26.9^*$ &  58.6  & 0.609 \\
             &This work &  13.5  &  0.434& 44.7 & 49.4 & 0.604 \\ \hline
1000 & Coppi92 &  3.83 & 0.154 & 21.4 &  24.2  & 0.749 \\
               &This work  & 3.29  &  0.234 & 28.0 &  20.4 & 0.702 \\ \hline
\end{tabular} 
\caption{Output parameters for synchrotron self-compton models. The magnetic field B is given in Gauss. Other parameters definitions are the same as in Table \ref{Coppi_tab1}. $^*$The pair yield and temperature given by \citet{Coppi92} for B=300G were multiplied by 10 to correct for what we believe to be typos.} \label{Coppi92_outputs2}
\end{center}
\end{table}
Although the general spectrum shape is recovered, substantial differences are observed in the far UV, soft X-ray band where the flux appears to have been underestimated in previous studies. We also derive a measurement of temperature for the thermal part of the distribution that is larger. This most likely results from our more precise treatment of the cyclotron emission/absorption. 

\subsection{The synchrotron boiler}
\begin{figure}[h!]
\begin{center}
\includegraphics[scale=.5]{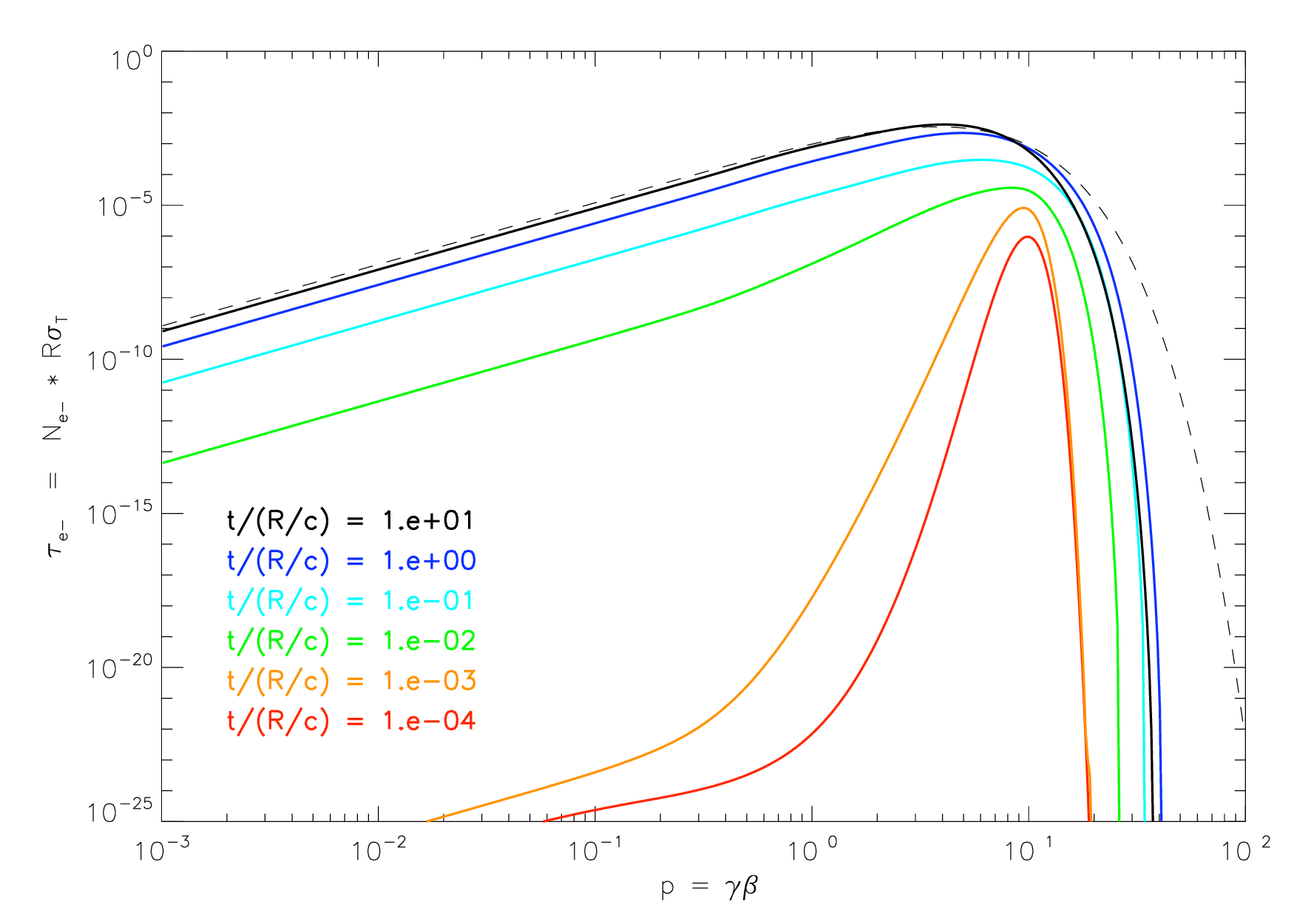}
\includegraphics[scale=.5]{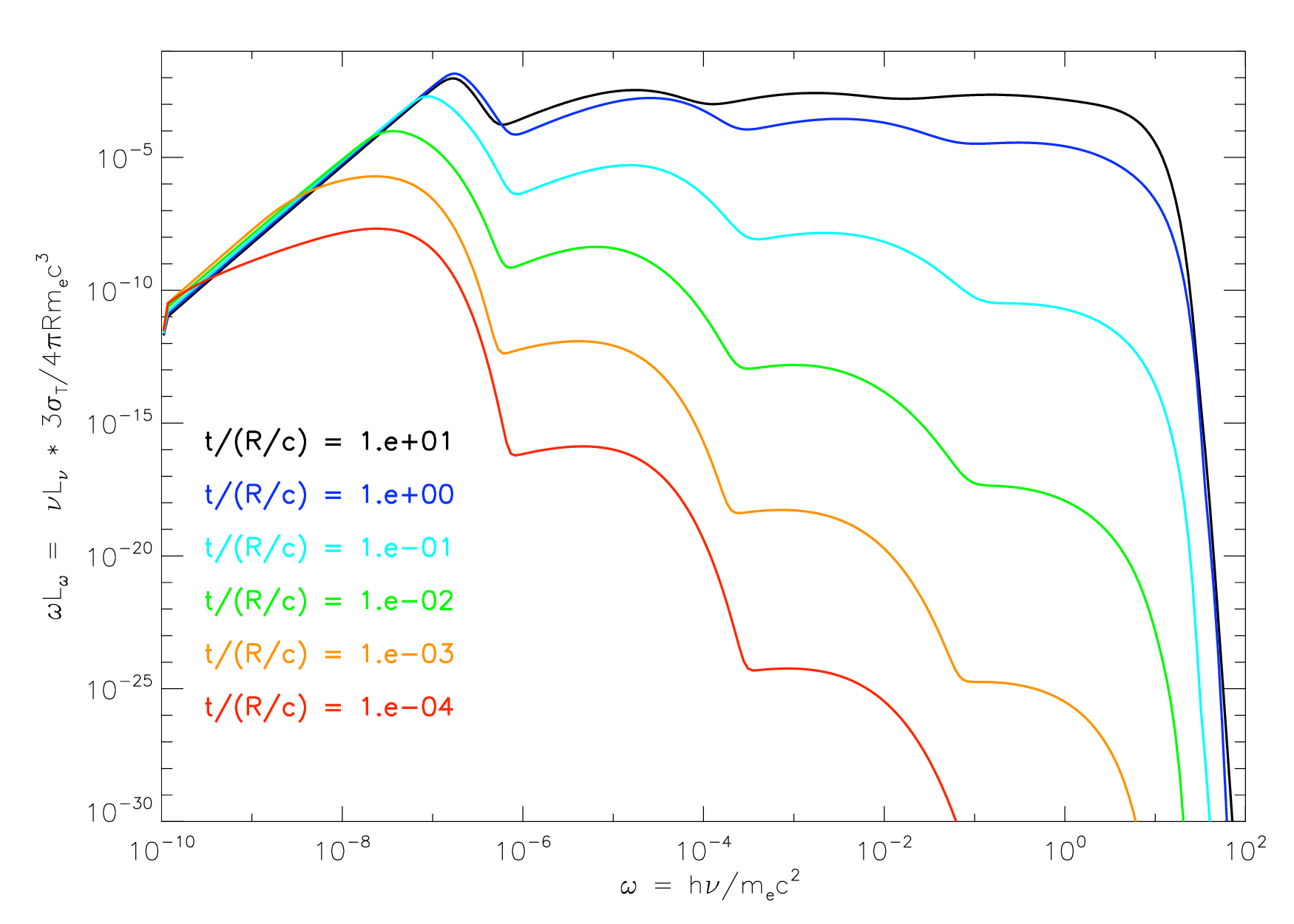}
\caption{Evolution of the particle distribution (upper panel) and the outgoing photon flux (lower panel). Times are $t/(R/c)=10^{-4}, 10^{-3}, 10^{-2}, 10^{-1}, 1$ and $10$ from lower to higher curves. Particles are injected with a Gaussian distribution centred at $\gamma=10$, of width $\delta \gamma =1$, and with a compactness parameter: $l_{e^-}=1$. The magnetic compactness is $l_B=10$ and the domain size is: $R=10^{13}$ cm. The dashed curve is the Maxwell-Boltzmann distribution of same normalization and same average energy as the equilibrium solution.} \label{GHS98a}
\end{center}
\end{figure}

Non-thermal distributions of particles can be thermalized by the emission and absorption of synchrotron photons. The efficiency of this mechanism however depends on the parameters. To illustrate this process, we consider the case presented in \citet{GHS98}, where high energy particles strongly emit and absorb synchrotron photons. We inject a constant mono-energetic distribution of electrons into an empty source of size $R=10^{13}$ cm. Electrons escape freely, which leads to a steady state. In this study, we consider only cyclo-synchrotron radiation and Compton scattering (pair effects, Coulomb scattering, and e-p bremsstrahlung are neglected). Figure \ref{GHS98a} shows the time evolution of the particle and photon distributions and a comparison with previous work. 

The results qualitatively confirm those of \citet{GHS98}. As time evolves, high energy particles are cooled by synchrotron radiation, which starts to create a radiation field. Soft synchrotron photons are up-scattered by high energy electrons and form the high energy part of the spectrum. For the choice of parameters ($l_{e^-}/l_B << 1$), the effect of synchrotron self-absorbed radiation on the particle distribution dominates over Compton scattering so that the additional cooling on particles by the latter is negligible. The synchrotron cooling timescale for particles is then $t_s/(R/c) = (\gamma-1)/(4l_Bp^2/3)\approx  1/(\gamma+1)/l_B$ and ranges between $0.5/l_B$ for low energy particles and $0.05/l_B$ for high energy particles ($\gamma\approx20$). For $l_B=10$, the distribution has reached a quasi-thermal shape at $t\approx 0.01-0.1 R/c$, i.e. on the synchrotron timescale (see Fig. \ref{GHS98a}). The normalization then saturates as the escape of particles balances the injection rate, which occurs on a typical time scale of $t_{esc} \gtrsim R/c$. The low energy part of the distribution is well reproduced by a Maxwell distribution but the high energy part of the distribution declines more rapidly than a real thermal distribution. In more detail, the results differ however quiet significantly. For the sake of clarity, the results of \citet{GHS98} have not been overplotted, athough their distributions are systematically colder and broader than the previous ones, especially in the transient phase. Although the low energy part still looks thermal, the deviation at high energy is therefore higher.

\begin{figure}[h!]
\begin{center}
\includegraphics[scale=.5]{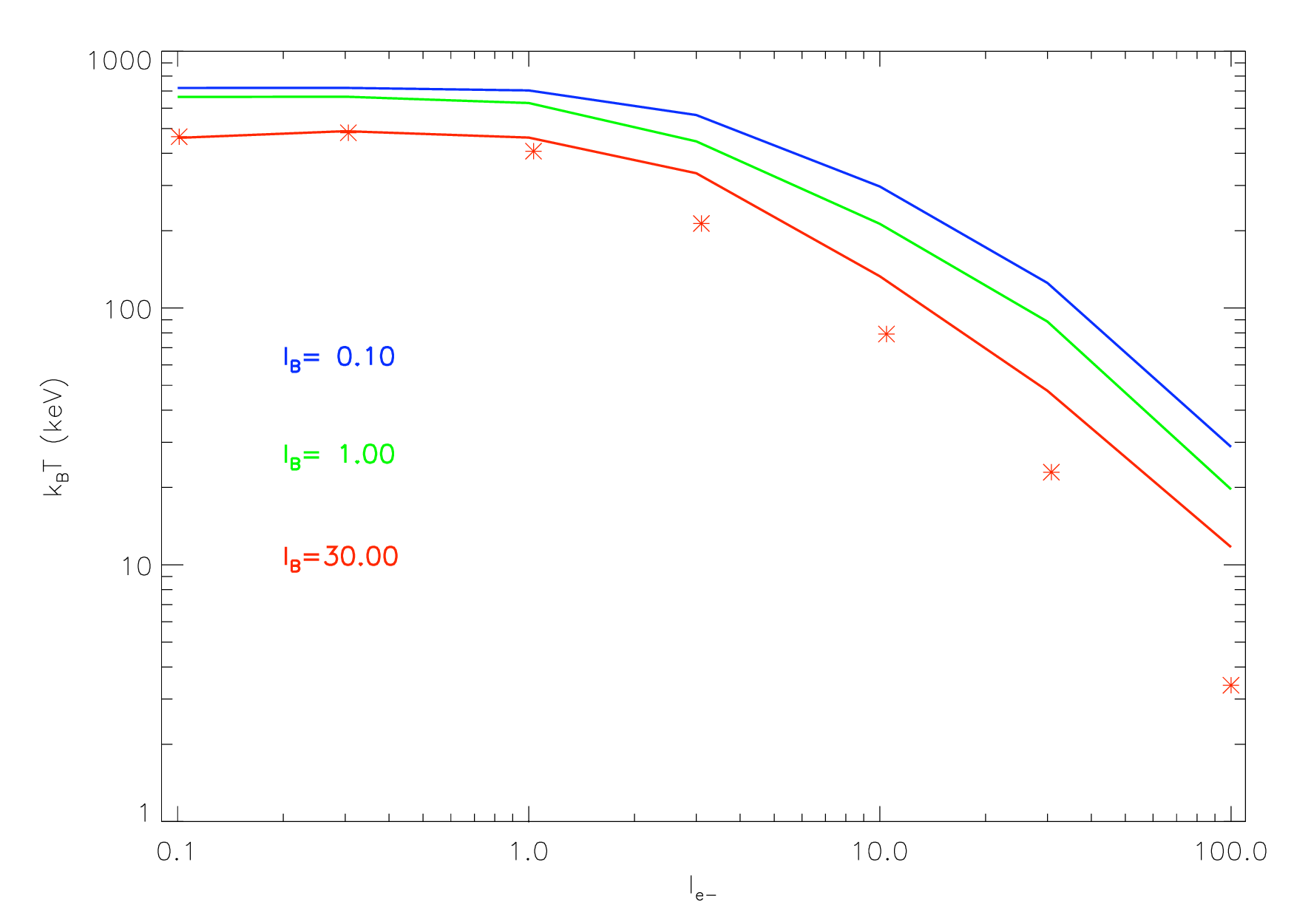}
\caption{Effective electron temperature estimated following Eq. (18) in \citet{GHS98}. The 3 curves are for magnetic compactness parameters $l_B=0.1,1,30$ from the top curve to the bottom curve and the domain size is $R=10^{13}$ cm. The stars indicate the results of \citet{GHS98} for $l_B=30$.} \label{GHS98C}
\end{center}
\end{figure}

To illustrate the effect of the magnetic field intensity and the particle injection rate, we plot in Fig. \ref{GHS98C} the temperature\footnote{the temperature is the effective temperature defined by Eq. (18) in \citet{GHS98}} of the steady distribution as a function of the injection compactness parameter for 3 different magnetic compactness parameters. 
The steady temperature is quite insensitive to the magnetic parameter since the temperature varies by only a factor of less than 3 as $l_B$ varies by over more than 2 orders of magnitude. In contrast, it is quite dependent on the injection compactness parameter. For a given magnetic field, the low injection rates produce steady states in which the Thomson optical depth is low. The Compton cooling is negligible and the final temperature is high. As the injection parameter increases, the optical depth becomes large, and the Compton cooling becomes efficient and dominates at high energy, eventually producing far smaller temperatures. The injection of particles at an energy far higher than the average particle temperature causes the formation of a hard non-thermal tail and a larger deviation from the Maxwell distribution at high energies. Compared to the results by \citet{GHS98}, we find temperatures that are significantly higher (up to a factor 3) at large optical depths. This is probably due to a more precise treatment of the radiation field and Compton scattering. \citet{GHS98} considered only the cooling of particles by inverse Compton scattering and assumed that it was limited in the Thomson regime. By using the exact Klein-Nishina cross section, we find more rapid photon escape and a weaker radiation field, whose cooling efficiency is lower.

\begin{figure}[h!]
\begin{center}
\includegraphics[scale=.5]{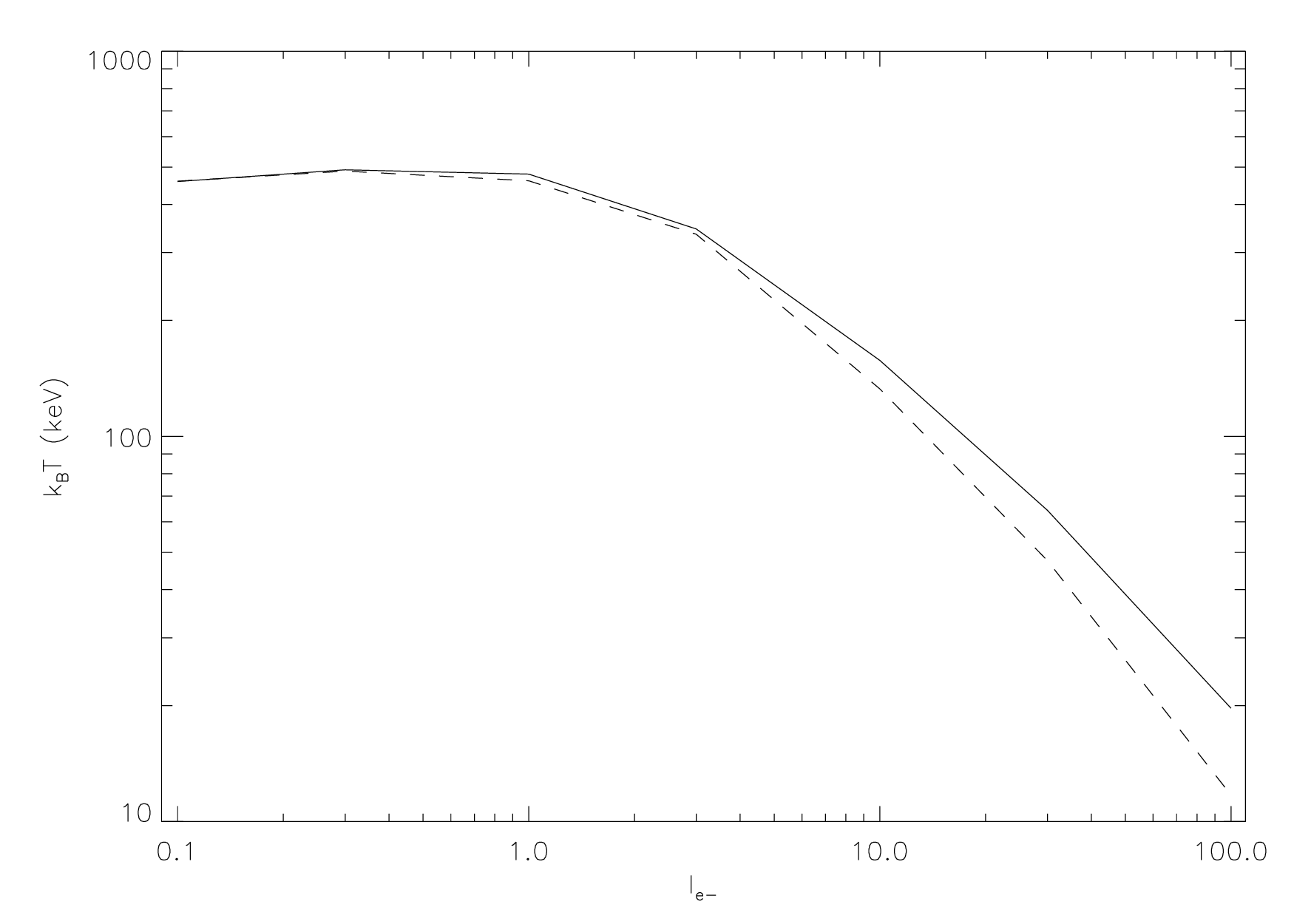}
\caption{Effect of Coulomb cooling: same as Fig. \ref{GHS98C} for $l_B=30$, with and without e-e Coulomb scattering  (solid and dashed line respectively). } \label{GHS98D}
\end{center}
\end{figure}

At large Thomson optical depth (i.e. at high injection rates), Coulomb exchange is supposed to dominate over synchrotron self-absorption. To investigate this, we completed the same simulations including e-e Coulomb scattering. Results are shown in Fig. \ref{GHS98D}.
It is found that the e-e Coulomb collisions tend to increase the effective temperature. As explained before, particles are injected at high energy. They are cooled by both synchrotron emission and Compton scattering, and form a low energy thermal pool. high energy particles are then scattered by thermal electrons with e-e Coulomb collisions. The cooling of the high energy distribution is very efficient but the thermal pool of cool electrons gain energy by this interaction, giving higher effective temperatures. This effect is negligible at low injection rates when the temperature is so high that the injection energy has a value that is almost in the bulk of the distribution and there is no well-marked high energy tail. However, at high injection rates, the temperature decreases and particles are injected at far higher energies than in the thermal pool. Exchange of energy between high and low energy particles becomes very efficient and it is found that this effect is significant (up to a factor of 2 for $l_e=100$).

A more detailed study of the synchrotron boiler mechanism and its application to X-ray binaries will be addressed in future work. 

\subsection{Particle acceleration}
As a second example, we investigate the effect of Fermi, second order acceleration. We consider a magnetised ($l_B=1$), isolated plasma of size $R=5\times10^7$ cm (typical of X-ray binary coronae), with no injection of seed photons. The soft photons are emitted by synchrotron radiation of high energy particles. The acceleration is modelled by the second order Fermi process and no particle is injected into the plasma. Particles are assumed to be trapped and the Thomson optical depth is set to be $\tau_e=1$. Pair production/annihilation and Coulomb collisions are neglected to focus on the role of particle acceleration. After a transient phase that depends on the initial conditions, particles and photons reach a steady state that depends only on the acceleration properties.

\begin{figure}[h!]
\begin{center}
\includegraphics[scale=.5]{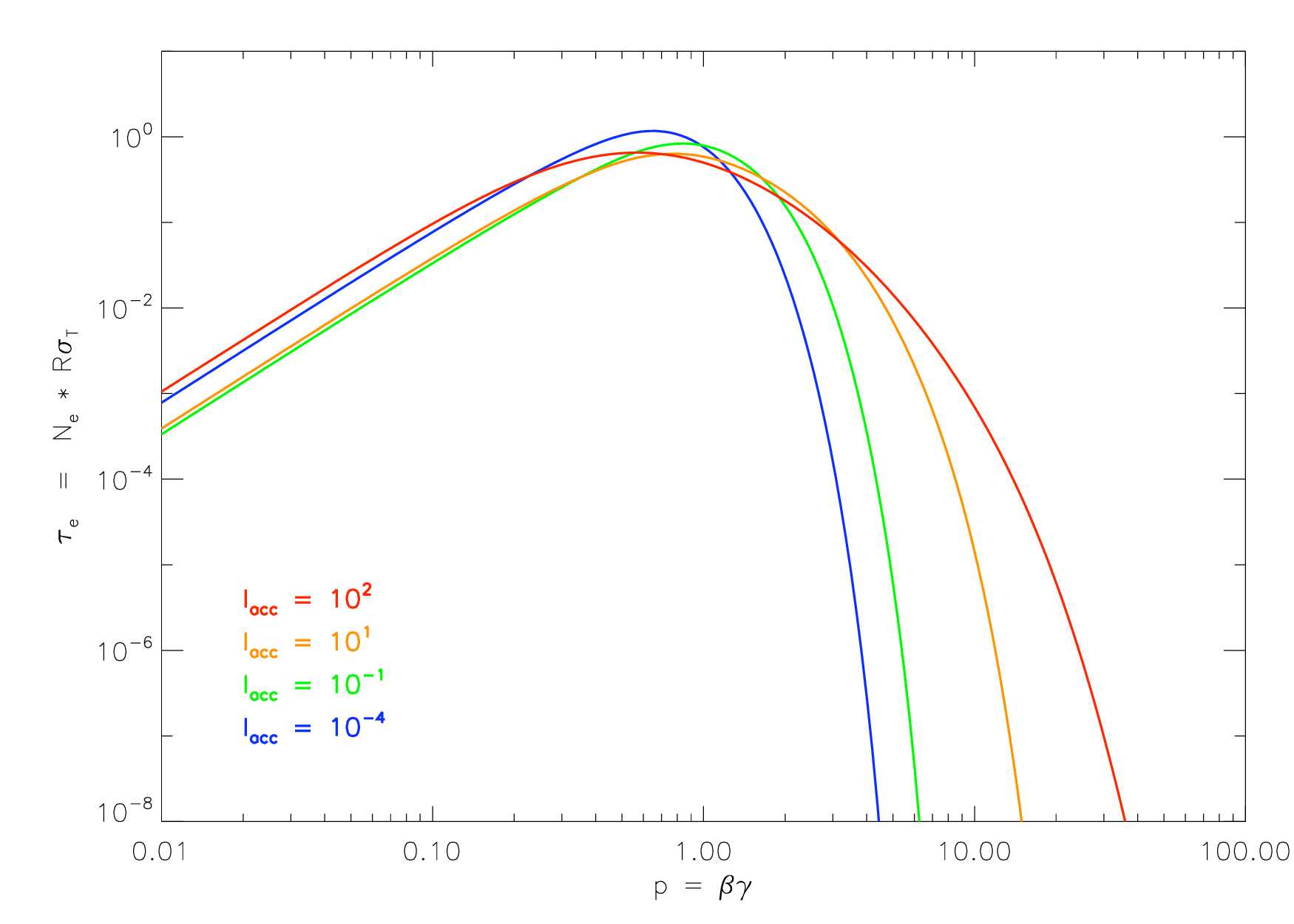}
\includegraphics[scale=.5]{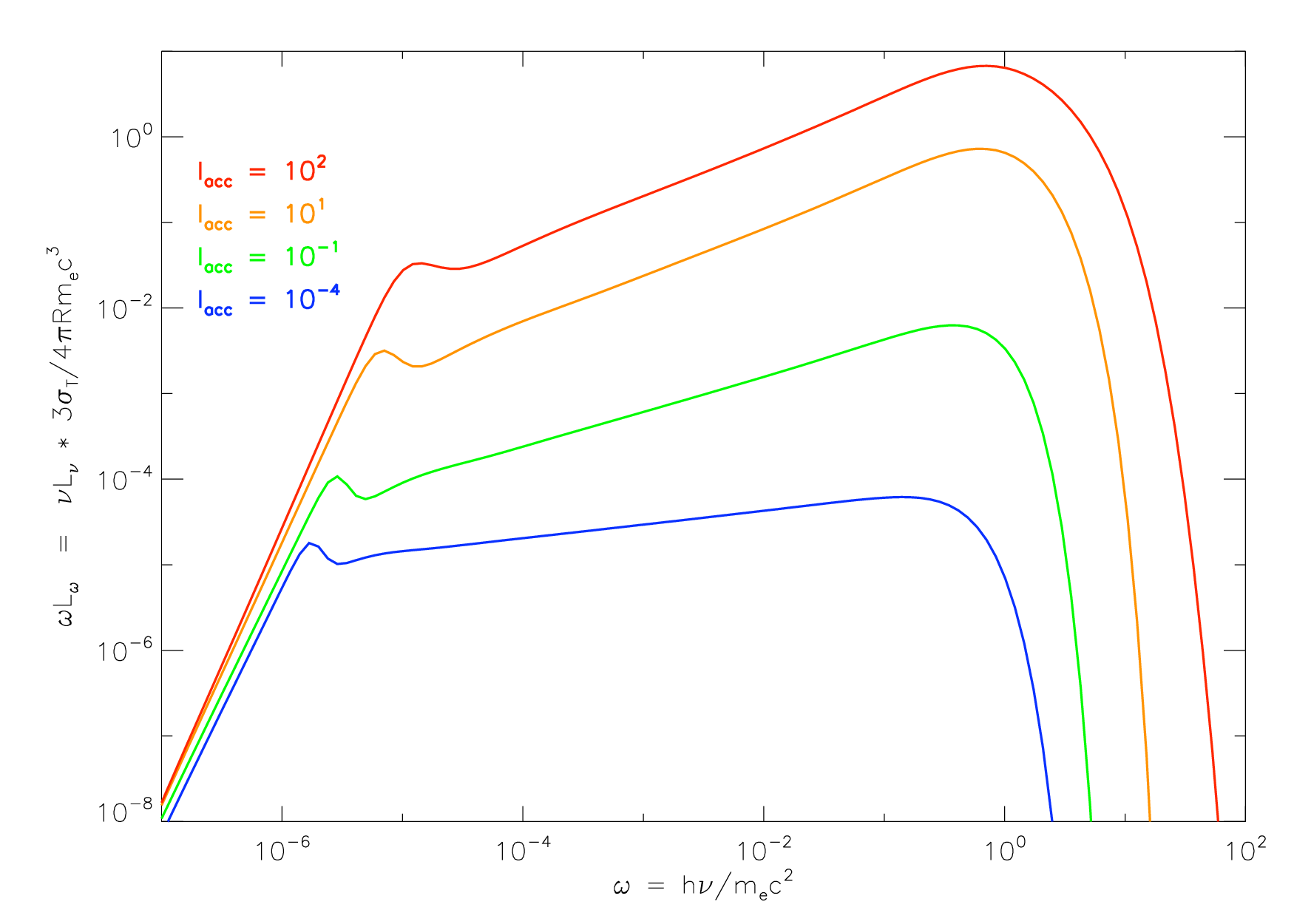}
\caption{Particle distributions (upper panel) and spectra (lower panel) for different acceleration efficiencies. Acceleration is modelled by second order Fermi process with no threshold. The optical depth is $\tau_e=1$, the domain size is $R=5\times 10^7$ cm and the magnetic compactness parameter is $l_B=1$.} \label{acc1}
\end{center}
\end{figure}

We first investigate the role of the acceleration efficiency and the threshold energy is assumed to be far lower than the bulk of particles. Figure \ref{acc1} presents the steady particle distributions and spectra for various values of the acceleration efficiency. 
In all cases, the distribution is similar to a Maxwell-Boltzmann distribution. As found in previous calculations, the diffusion in the momentum space produces a quasi-thermal distribution \citep[e.g.][]{Katar06} and in this case, the thermalisation is also helped by the synchrotron boiler mechanism. The spectrum is the sum of the low energy synchrotron emission and a hard tail resulting from the multiple Compton scattering of these soft photons from the highest energy particles. As the acceleration efficiency increases the steady distribution widens and moves to higher energies. As a consequence, the spectra exhibit a stronger hard tail. The temperature of the distribution is given in Table \ref{acc1_table}.
\begin{table}[h!]
\begin{center}
\begin{tabular}{c|cc|}
log($l_{\rm{acc}}$) & $\bar{t}_{\rm{acc}}$ & $\theta_3$  \\  \hline 
-4  & $48800 $  & 164.5   \\
-1  &  75.1 & 259.8   \\
1 & 1.02 &   369.0  \\
2 &  0.104 &  390.5  \\
 \hline
\end{tabular}
\caption{Outputs parameters for Fig. \ref{acc1}. $\bar{t}_{\rm{acc}}$ is the acceleration time in unit $R/c$ and $\theta_3$ is the averaged temperature estimated as in section \ref{coppi_sec} in units of $10^{-3} m_ec^2$. } \label{acc1_table}
\end{center}
\end{table}
and as expected, it becomes higher as the acceleration efficiency increases. Since there are more high energy particles, the synchrotron self-absorbed emission is stronger, which cools the softer particles more efficiently and the averaged temperature saturates. The source luminosity however scales with the acceleration compactness parameter.  

\begin{figure}[h!]
\begin{center}
\includegraphics[scale=.5]{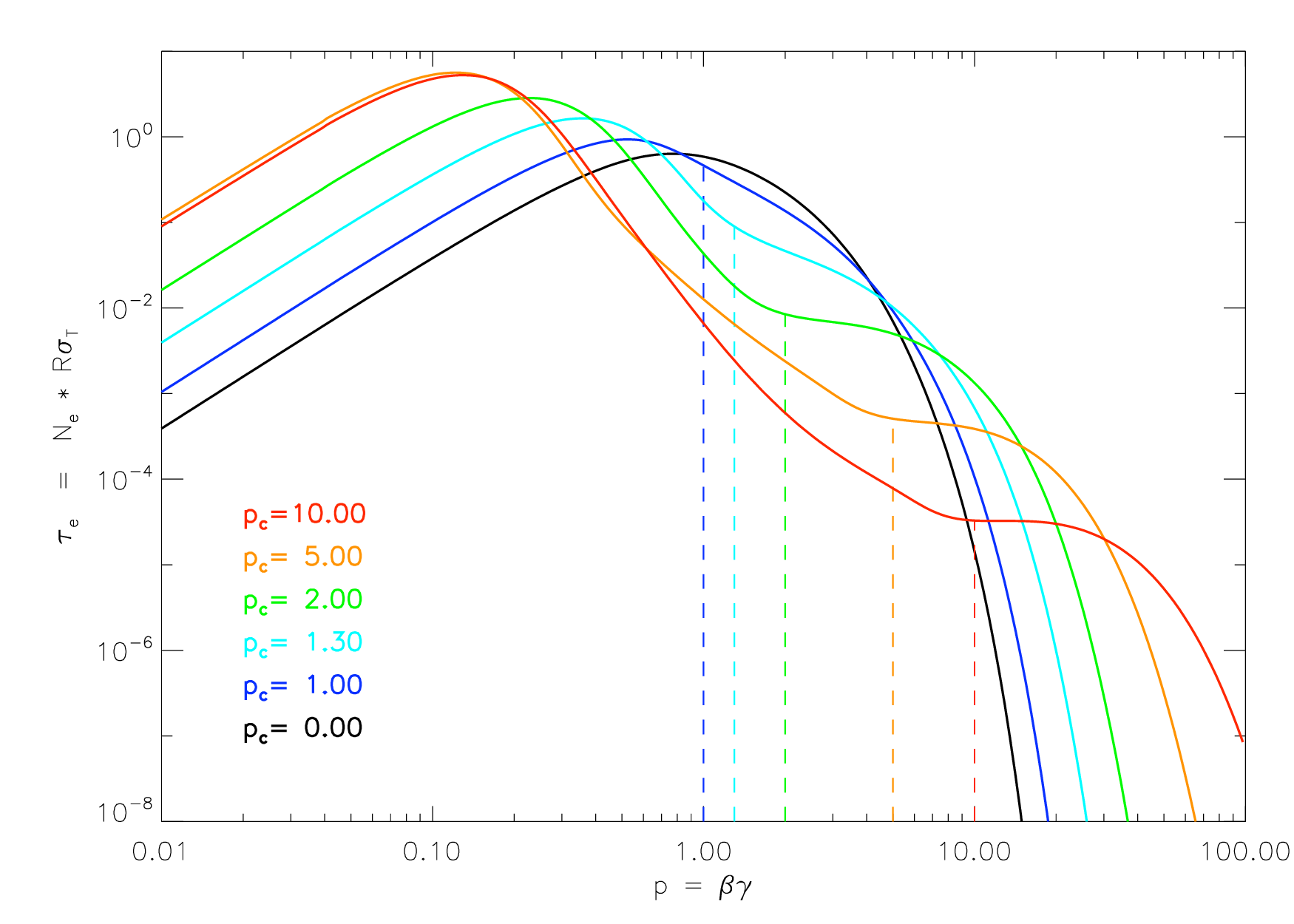}
\includegraphics[scale=.5]{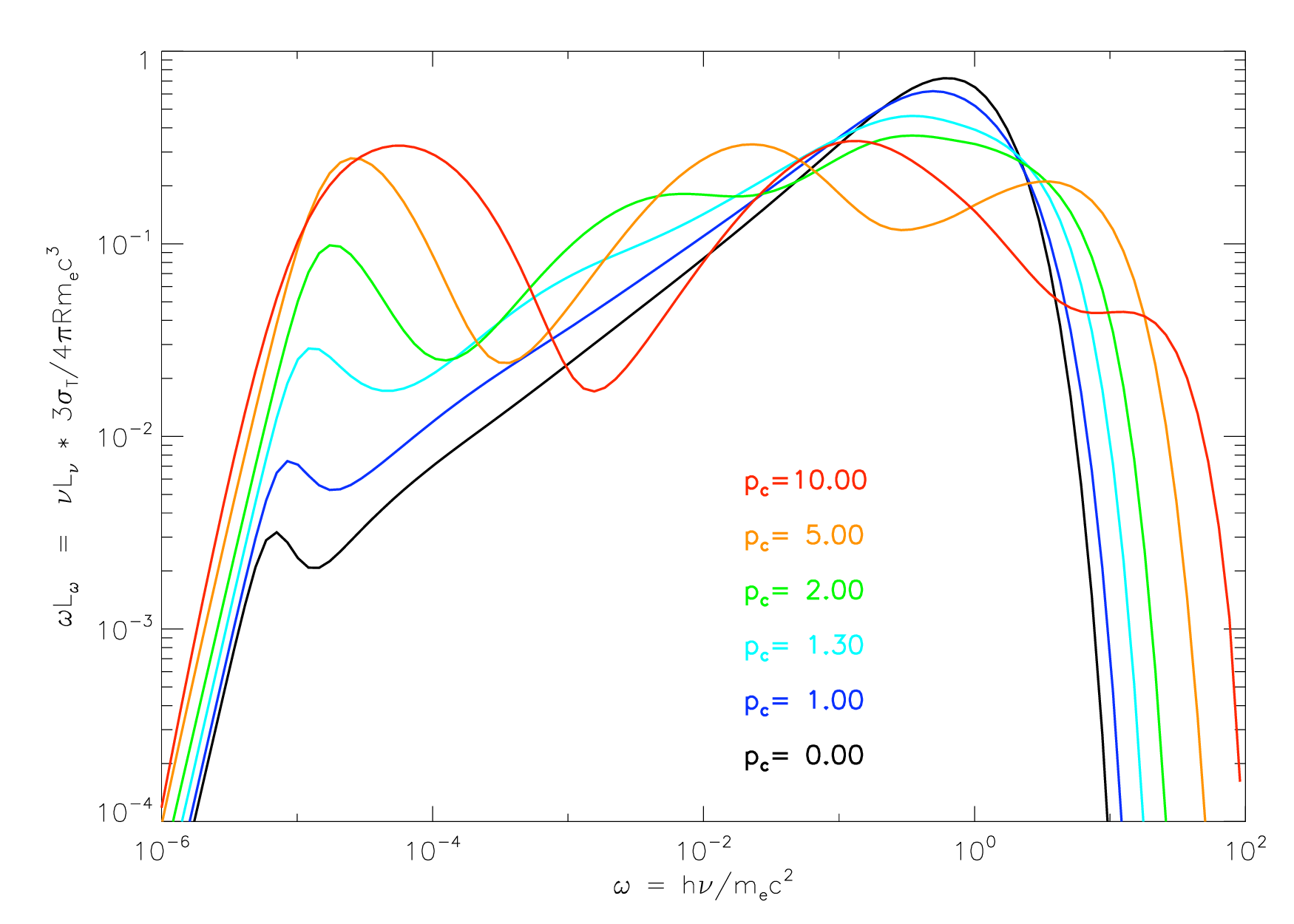}
\caption{Particle distributions (upper panel) and spectra (lower panel) for different threshold energies. Acceleration is modelled by second order Fermi process with $l_{\rm{acc}}=10$. The optical depth is $\tau_e=1$, the domain size is $R=5\times 10^7$ cm and the magnetic compactness parameter is $l_B=1$. Dashed vertical lines show the threshold momenta. } \label{acc2}
\end{center}
\end{figure}

We now investigate the role of the minimal energy above which the acceleration occurs. The simulations are completed with the same parameters but the acceleration efficiency is set to $l_{\rm{acc}}=10$ and we vary the threshold energy. We note that for a given power supplied to the plasma by accelerating particles, the higher the threshold energy the fewer the accelerated particles and the shorter their acceleration time. The results are shown in Fig. \ref{acc2} for threshold momenta $p_c=0,1,1.3,2,5$, and 10.
The corresponding acceleration times are $t_{\rm{acc}}=1.02,6.32\times10^{-1},3.45\times10^{-1},1.77\times10^{-1},5.63\times10^{-2},$ and $1.95\times10^{-2} R/c$ respectively.
For low threshold energies ($p_c \le 1$), the distribution depends weakly on the precise threshold. It produces a small number of soft photons. Since the particles are not energetic, the photons undergo multiple Compton scattering and form the strong high energy part of the spectrum. When only mid-relativistic or relativistic particles are accelerated, they form a high energy tail that extends far beyond the thermal pool, and the situation then differs significantly. Since the total synchrotron emission increases with the energy of the emitting particles ($\int j_s d\nu \propto p^2$), the synchrotron bump is much larger. The hard energy tail of the particle distribution also has a flat slope, which produces a wider synchrotron bump. Since the particles have higher energy, the Compton up-scattering of these soft photons becomes more efficient. In the limit where the synchrotron soft photons have low energy ($\omega_{\rm{synch}}  << 1$) and the accelerated electrons have high energy ($\gamma >> 1$), the photon energy gain during one single scattering is: $\omega_{\rm{compt}}/\omega_{\rm{synch}}\approx 4\gamma^2/3$. As a result, photons undergo only a small number of Compton scatterings before they reach the particle energy, forming a double-humped spectrum as those of blazars in the case $p_c=10$. At the same time, Compton scattering cools the thermal particles further so that the bulk of particles moves to lower energies. Future work will consider this effect in more details including the physics of wave-particle interaction. In particular, more consistent models of particle escape and momentum diffusion must be implemented.

\section*{Conclusion}
We have presented a code developed to model radiation processes in high energy plasmas without any assumption about the shape of the particle distribution. The code is time dependent. It uses the exact Compton and pair production/annihilation unpolarized, isotropic cross sections. Cyclo-synchrotron self-absorbed radiation is taken into account from the sub-relativistic regime to the ultra-relativistic one, which represent an improvement in comparison with other codes. It also includes an approximate treatment of e-e and e-p Coulomb exchange and e-p self-absorbed bremsstrahlung radiation. Explicit prescriptions for particle acceleration have also been implemented. The code deals consistently with all processes over wide ranges of energy. There is no restriction on the photon energy and particles can have momenta in the range $10^{-7} \lesssim p \lesssim 10^{7}$. It can therefore be used to model various sources, such as not only X-ray binaries and AGN, but also $\gamma$-ray bursts and pulsar wind nebulae. 
 
The major limitation of the code is its simplified geometry. The code simulates a uniform system, typically a homogeneous sphere with an isotropic and unpolarized radiation field. It obviously introduces a bias in simulations of X-ray binaries coronae, where the seed photons from the disc have an isotropic distribution or in jets of AGN, where geometrical effects are important. However, the geometry of the emitting regions in high energy sources is poorly constrained and in most cases it does not play a crucial role. The prescriptions used for particle and photon escape are also able to reproduce the main effects of geometry. 

Some examples have been shown of checks of the code capabilities in comparison with previous codes designed to solve restricted problems, involving a limited number of ingredients. In several cases, we have disabled some processes in our code to ensure more rigourous comparisons. We have found that the code confirms qualitatively all previous results. After considering more precisely these processes, the properties of the exact spectra and particle distribution were however found to be slightly different. As an example, we investigated the acceleration by second order Fermi-like processes. We found that an energy threshold for acceleration produces a non-thermal population of particles when it reaches the mid-relativistic regime.
 

\bibliographystyle{aa}

\begin{thebibliography}{}
\bibitem[Alexanian(1968)]{Alexanian68} Alexanian, M.\ 1968, Physical Review , 165, 253
\bibitem[Belmont(2008)]{Belmont08} Belmont, R., \aap , 2008, submitted to A\&A
\bibitem[Bisnovatyi-Kogan et al.(1971)]{BZS71} Bisnovatyi-Kogan, G.~S., Zel'Dovich, Y.~B., \& Syunyaev, R.~A.\ 1971, Soviet Astronomy, 15, 17
\bibitem[Blasi(2000)]{Blasi00} Blasi, P.\ 2000, \apjl, 532, L9
\bibitem[Brinkmann(1984)]{Brinkmann84} Brinkmann, W.\ 1984, Journal of Quantitative Spectroscopy and Radiative Transfer, 31, 417
\bibitem[Boettcher \& Schlickeiser(1997)]{BS97} Boettcher, M., \& Schlickeiser, R.\ 1997, \aap, 325, 866
\bibitem[B{\"o}ttcher \& Liang(2001)]{BL01} B{\"o}ttcher, M., \& Liang, E.~P.\ 2001, \apj, 552, 248 
\bibitem[Chang \& Cooper(1970)]{CC70} Chang, J.~S., \& Cooper, G.\ 1970, Journal of Computational Physics, 6, 1
\bibitem[Coppi(1992)]{Coppi92} Coppi, P.~S.\ 1992, \mnras, 258, 657
\bibitem[Courant, Friedrichs \& Lewy(1928)]{CFL} R. Courant, K. Friedrichs \& Lewy, H. \ 1967, IBM Journal, 215, English translation of the 1928 German original
\bibitem[Crusius \& Schlickeiser(1986)]{CS86} Crusius, A., \& Schlickeiser, R.\ 1986, \aap, 164, L16 
\bibitem[Dermer \& Liang(1989)]{DL89} Dermer, C.~D., \& Liang, E.~P.\ 1989, \apj, 339, 512
\bibitem[Dermer et al.(1996)]{DML96} Dermer, C.~D., Miller, J.~A., \& Li, H.\ 1996, \apj, 456, 106
\bibitem[Fabian et al.(1986)]{FBGPC86} Fabian, A.~C., Guilbert, P.~W., Blandford, R.~D., Phinney, E.~S., \& Cuellar, L.\ 1986, \mnras, 221, 931
\bibitem[Ghisellini(1987)]{Ghisellini87} Ghisellini, G.\ 1987, \mnras, 224, 1 
\bibitem[Ghisellini et al.(1988)]{GGS88} Ghisellini, G., Guilbert, P.~W., \& Svensson, R.\ 1988, \apjl, 334, L5
\bibitem[Ghisellini \& Svensson(1991)]{GS91} Ghisellini, G., \& Svensson, R.\ 1991, \mnras, 252, 313
\bibitem[Ghisellini et al.(1993)]{GHF93} Ghisellini, G., Haardt, F., \& Fabian, A.~C.\ 1993, \mnras, 263, L9
\bibitem[Ghisellini et al.(1998a)]{GHS98} Ghisellini, G., Haardt, F., \& Svensson, R.\ 1998, \mnras, 297, 348
\bibitem[Ghisellini et al.(1998b)]{Ghisellini98} Ghisellini, G., Celotti, A., Fossati, G., Maraschi, L., \& Comastri, A.\ 1998, \mnras, 301, 451
\bibitem[Gorecki \& Wilczewski(1984)]{Gorecki84} Gorecki, A., \& Wilczewski, W.\ 1984, Acta Astronomica, 34, 141
\bibitem[Gould(1980)]{Gould80} Gould, R.~J.\ 1980, \apj, 238, 1026
\bibitem[Guilbert(1981)]{Guilbert81} Guilbert, P.~W.\ 1981, \mnras, 197, 451 
\bibitem[Guilbert \& Stepney(1985)]{Guilbert85} Guilbert, P.~W., \& Stepney, S.\ 1985, \mnras, 212, 523 
\bibitem[Guilbert(1986)]{Guilbert86} Guilbert, P.~W.\ 1986, \mnras, 218, 171
\bibitem[Haug(1975)]{Haug75} Haug, E.\ 1975, Zeitschrift Naturforschung Teil A, 30, 1099
\bibitem[Haug(1985)]{Haug85} Haug, E.\ 1985, \prd, 31, 2120
\bibitem[Heitler(1954)]{Heitler54} Heitler, W.\ 1954, International Series of Monographs on Physics, Oxford: Clarendon, 1954, 3rd ed.
\bibitem[Jauch \& Rohrlich(1976)]{JR76} Jauch, J.~M., \& Rohrlich, F.\ 1976, Texts and Monographs in Physics, New York: Springer, 1976, 2nd ed.
\bibitem[Jones(1968)]{Jones68} Jones, F.~C.\ 1968, Physical Review , 167, 1159
\bibitem[Katarzy{\'n}ski et al.(2006a)]{KGSG06} Katarzy{\'n}ski, K., Ghisellini, G., Svensson, R., \& Gracia, J.\ 2006, \aap, 451, 739
\bibitem[Katarzy{\'n}ski et al.(2006b)]{Katar06} Katarzy{\'n}ski, K., Ghisellini, G., Mastichiadis, A., Tavecchio, F., \& Maraschi, L.\ 2006, \aap, 453, 47 
\bibitem[Kusunose(1987)]{Kusunose87} Kusunose, M.\ 1987, \apj, 321, 186 
\bibitem[Le Roux(1960)]{LeRoux60} Le Roux, E.\ 1960, Annales d'Astrophysique, 23, 1010
\bibitem[Li et al.(1996)]{LKL96} Li, H., Kusunose, M., \& Liang, E.~P.\ 1996, \apjl, 460, L29
\bibitem[Li \& Miller(1997)]{Li97} Li, H., \& Miller, J.~A.\ 1997, \apjl, 478, L67 
\bibitem[Lightman et al.(1987)]{LZR87} Lightman, A.~P., Zdziarski, A.~A., \& Rees, M.~J.\ 1987, \apjl, 315, L113
\bibitem[Lightman \& Zdziarski(1987)]{LZ87} Lightman, A.~P., \& Zdziarski, A.~A.\ 1987, \apj, 319, 643
\bibitem[Le Roux(1961)]{Leroux61} Le Roux, E.\ 1961, Annales d'Astrophysique, 24, 71
\bibitem[McCray(1969)]{McCray69} McCray, R.\ 1969, \apj, 156, 329
\bibitem[Malzac \& Jourdain(2000)]{mj00} Malzac, J., \& Jourdain, E.\ 2000, \aap, 359, 843 
\bibitem[Marcowith \& Malzac(2003)]{Marcowith03} Marcowith, A., \& Malzac, J.\ 2003, \aap, 409, 9
\bibitem[Nagirner \& Poutanen(1994)]{NP94} Nagirner, D.~I., \& Poutanen, J.\ 1994, Single Compton scattering, Astrophysics and Space Physics Reviews,   vol.~9, part 1.~ Amsterdam: Harwood Academic Publishers,  c1994, 83 pages.,
\bibitem[Nayakshin \& Melia(1998)]{NM98} Nayakshin, S., \& Melia, F.\ 1998, \apjs, 114, 269
\bibitem[Park \& Petrosian(1996)]{PP96} Park, B.~T., \& Petrosian, V.\ 1996, \apjs, 103, 255
\bibitem[Pe'er \& Waxman(2005)]{PW05} Pe'er, A., \& Waxman, E.\ 2005, \apj, 628, 857 
\bibitem[Poutanen \& Svensson(1996)]{Poutanen96} Poutanen, J., \& Svensson, R.\ 1996, \apj, 470, 249
\bibitem[Pozdnyakov et al.(1977)]{PSS77} Pozdnyakov, L.~A., Sobol, I.~M., \& Siuniaev, R.~A.\ 1977, Soviet Astronomy, 21, 708
\bibitem[Pozdnyakov et al.(1980)]{PSS80} Pozdnyakov, L.~A., Sobol, I.~M., \& Syunyaev, R.~A.\ 1980, Comptomization and radiation spectra of X-ray sources.~ Calculation of the Monte Carlo method,  Rept.~Pr-447 Acad.~of Sci.~USSR, Moscow, 1978  12 p,
\bibitem[Pozdnyakov et al.(1983)]{PSS83} Pozdnyakov, L.~A., Sobol, I.~M., \& Siuniaev, R.~A.\ 1983, Astrophysics and Space Physics Reviews, 2, 189
\bibitem[Stepney(1983)]{Stepney83a} Stepney, S.\ 1983, \mnras, 202, 467
\bibitem[Stepney \& Guilbert(1983)]{Stepney83b} Stepney, S., \& Guilbert, P.~W.\ 1983, \mnras, 204, 1269
\bibitem[Stern et al.(1995)]{Stern95} Stern, B.~E., Begelman, M.~C., Sikora, M., \& Svensson, R.\ 1995, \mnras, 272, 291
\bibitem[Sunyaev \& Titarchuk(1980)]{Sunyaev80} Sunyaev, R.~A., \& Titarchuk, L.~G.\ 1980, \aap, 86, 121
\bibitem[Svensson(1982a)]{Svensson82a} Svensson, R.\ 1982, \apj, 258, 321
\bibitem[Svensson(1982b)]{Svensson82b} Svensson, R.\ 1982, \apj, 258, 335
\bibitem[Svensson(1983)]{Svensson83} Svensson, R.\ 1983, \apj, 270, 300
\bibitem[Svensson(1984)]{Svensson84} Svensson, R.\ 1984, \mnras, 209, 175
\bibitem[Svensson(1987)]{Svensson87} Svensson, R.\ 1987, \mnras, 227, 403
\bibitem[Wardzi{\'n}ski \& Zdziarski(2000)]{Wardzinski00} Wardzi{\'n}ski, G., \& Zdziarski, A.~A.\ 2000, \mnras, 314, 183
\bibitem[Zdziarski(1984)]{Zdziarski84} Zdziarski, A.~A.\ 1984, \apj, 283, 842
\bibitem[Zdziarski(1985)]{Zdziarski85} Zdziarski, A.~A.\ 1985, \apj, 289, 514
\bibitem[Zdziarski et al.(1993)]{ZLM93} Zdziarski, A.~A., Lightman, A.~P., \& Maciolek-Niedzwiecki, A.\ 1993, \apjl, 414, L93
\end{thebibliography}


\end{document}